\documentclass[AMA,LATO1COL]{WileyNJD-v2}
\usepackage{xcolor}
\usepackage{subfig}
\usepackage{hyperref}

\articletype{Research Article}

\received{}
\revised{}
\accepted{}

\begin{document}

\title{Evaluating the Potential of Disaggregated Memory Systems for HPC applications}

\author[1]{Nan Ding*}
\author[3]{Pieter Maris}
\author[2]{Hai Ah Nam}
\author[2]{Taylor Groves}
\author[2]{Muaaz Gul Awan}
\author[2]{LeAnn Lindsey}
\author[2]{Christopher Daley}
\author[1]{Oguz Selvitopi}
\author[1]{Leonid Oliker}
\author[2]{Nicholas Wright}
\author[1]{Samuel Williams}

\authormark{Nan Ding. \textsc{et al}}

\address[1]{\orgdiv{Computational Research Division}, \orgname{Lawrence Berkeley National Laboratory}, 
\orgaddress{\state{CA}, \country{USA}}}

\address[2]{\orgdiv{National Energy Research Scientific Computing Center}, \orgname{Lawrence Berkeley National Laboratory}, \orgaddress{\state{CA}, \country{USA}}}

\address[3]{\orgdiv{Department Of Physics and Astronomy}, \orgname{Iowa State University}, 
\orgaddress{\state{IA}, \country{USA}}}

\corres{*Nan Ding \email{nanding@lbl.gov}}

\presentaddress{1 Cyclotron Road, Berkeley, CA 94720, USA}

\authormark{Nan Ding \textsc{et al}}

\abstract[Summary]{
Disaggregated memory is a promising approach that addresses the limitations of traditional memory architectures by enabling memory to be decoupled from compute nodes and shared across a data center. Cloud platforms have deployed such systems to improve overall system memory utilization, but performance can vary across workloads. High-performance computing (HPC) is crucial in scientific and engineering applications, where HPC machines also face the issue of underutilized memory. As a result, improving system memory utilization while understanding workload performance is essential for HPC operators. Therefore, learning the potential of a disaggregated memory system before deployment is a critical step. This paper proposes a methodology for exploring the design space of a disaggregated memory system. It incorporates key metrics that affect performance on disaggregated memory systems: memory capacity, local and remote memory access ratio, injection bandwidth, and bisection bandwidth, providing an intuitive approach to guide machine configurations based on technology trends and workload characteristics. We apply our methodology to analyze thirteen diverse workloads, including AI training, data analysis, genomics, protein, fusion, atomic nuclei, and traditional HPC bookends. Our methodology demonstrates the ability to comprehend the potential and pitfalls of a disaggregated memory system and provides motivation for machine configurations. Our results show that eleven of our thirteen applications can leverage injection bandwidth disaggregated memory without affecting performance, while one pays a rack bisection bandwidth penalty and two pay the system-wide bisection bandwidth penalty. 
In addition, we also show that intra-rack memory disaggregation would meet the application's memory requirement and provide enough remote memory bandwidth.
}

\keywords{Memory disaggregation, system design space, methodology for system design}

\maketitle


\section{Introduction}\label{sec:intro}
As disaggregated memory systems become increasingly practical and performant for deployment in the cloud~\cite{li2023pond}, they have garnered attention as a solution to improve memory utilization while reducing costs for High-Performance Computing (HPC) systems. Over the decades, HPC system architects have been forced to overprovision systems to meet long-tail memory requirements, deploying various node architectures that lead to inflexible resource usage or demanding that applications restructure themselves to fit within a smaller memory footprint. This results in computing nodes whose resource utilization can vary greatly from one application to another within a workload. For instance, NERSC's Cori supercomputer observed that only 15\% of scientific workloads utilize over 75\% of available memory per node~\cite{austin2020}. At Lawrence Livermore National Laboratory, 90\% of jobs use less than 15\% of node memory capacity~\cite{peng2020memory}. Furthermore, up to 83\% of memory can be underutilized on tightly-coupled resources that are over-provisioned for workloads with the highest demands~\cite{michelogiannakis2022case}.

Recent improvements in interconnect technology~\cite{li2022first} have reinvigorated memory disaggregation as a viable solution to both the memory and file system stranded resource problems, as many applications and workflows leverage high-performance distributed file systems for rather mundane tasks --- holding read-only or private files --- resulting in an overprovisioning of file system performance and degrading QoS for the applications that truly need a high-performance distributed file system.
Memory disaggregation decouples compute and memory resources. Compute nodes would contain only a limited amount of local memory, but could access a large pool of remote memory available via the network. This design enables HPC systems to scale memory capacity and allocate memory more flexibly. 
Physically, this large pool of memory would be partitioned among several smaller ``memory nodes'' each containing DRAM and a NIC in order to maximize bandwidth, capacity, and reliability.

Computer architects have continuously added more levels of caching to the memory hierarchy to bridge the performance gap for applications with significant temporal and spatial locality. Even modern GPU-accelerated systems have a hierarchy of faster and smaller memories, including CPU-attached DDR high-capacity memory, GPU-attached HBM high-performance memory, and multiple levels of on-GPU SRAM cache memories. Such design patterns will persist in systems like NVIDIA's Hopper H100 GPU~\cite{elster2022nvidia} but in a more tightly integrated and efficient form. Ultimately, this will enable single-chip CPU-GPU architectures called Accelerated Processing Units (APUs).

This work substantially expands our previous work~\cite{ding2022methodology}, which developed a methodology for modeling the performance implications of adding disaggregated memory to an APU-only HPC system. 
Whereas the previous paper bounded performance of 11 applications using only local (HBM) and network injection bandwidth, in this paper, we add two memory-intensive applications and examine the performance impact from the projected intra-rack and system-side bisection bandwidth limits of future HPC systems.
The new contributions of this paper include:

\begin{itemize}
\item Expanding our methodology to incorporate the effects of bisection bandwidth on disaggregated memory performance. The bisection bandwidth is often used for an upper bound on collective communications, e.g., all-to-all. Its impact can be amplified on a disaggregated memory system because users can load data from anywhere in the entire memory pool, and both inter-process communication and loading data from remote memory contend for the available memory node NIC bandwidth.
\item Addition of the direct sparse linear solver SuperLU\_DIST from M3CD1 ~\cite{Jardin2008M3dc1} (Fusion) and an iterative eigensolver from MFDn~\cite{maris2022accelerating} (atomic nucle) to the eleven applications in our original paper. These two workloads represent two different traditional HPC workloads: sparse triangular solves with sparse LU factorization in SuperLU\_DIST and sparse matrix-matrix multiply in eigensolver.
\item Release of profiling and plotting capabilities\cite{dis_mem_script} including system architecture design space, application performance potentials, and profiling scripts. Readers should be able to apply our methodology and analysis to their own kernels or applications with the scripts and Python 3.10.
\end{itemize}

\section{Related Work}\label{sec:relwork}
The growing interest and maturity of Compute Express Link (CXL)~\cite{cxl, cxl2, pcie,van2019hoti}, a standardized protocol for memory pooling, has been contributing to this renewed interest in memory disaggregation. CXL provides memory coherency and semantics over the PCIe physical layer. Pond~\cite{li2023pond, li2022first}, the first full-stack disaggregated memory system using CXL in cloud platforms, shows that it can reduce DRAM needs by 7\% with a memory pool of 16 sockets, which corresponds to hundreds of millions of dollars for a large cloud provider. The study also shows that 21\% of 158 workloads have more than a 25\% slowdown.

Utilization analysis for HPC systems report the average memory utilization of a job can be as small as 11.9\% and 74.6\% of individual jobs never use more than 50\% of on-node memory. Approximately three-quarters of the time, each compute node uses only 0.3\% of memory bandwidth and 0.5\% of available NIC bandwidth~\cite{michelogiannakis2022case}. Often resources are idle, since HPC system node design is based on the peak usage, i.e., the maximum memory usage. It is worth mentioning that DRAM consumes static power even when idle, so unused memory still contributes to the HPC system operating cost~\cite{peng2020memory}. Over the years, it has become a common state of practice for memory resources on HPC systems to be over-provisioned and have on-average low utilization.

Recent advancements in interconnect technology have led to the proposal of Network-Attached Disaggregated Memory for HPC systems~\cite{lim2009disaggregated,gu2017efficient}. Peng et al. further designed a user-space remote paging library to allow applications to explore the potential of throughput scaling on disaggregated memory~\cite{peng2020memory}. Jacob el al. developed an emulator to evaluate a memory subsystem design leveraging CXL-enabled memory pooling and demonstrated that a disaggregated memory system can effectively support bandwidth-intensive unstructured mesh-based applications like OpenFOAM~\cite{wahlgren2022evaluating}. Debendra discussed the potential and limitations of using CXL to build composable and scale-out systems spanning the rack through the pod at the data center~\cite{sharma2023novel}.
Furthermore, many studies focus on efficient memory management~\cite{maruf2022tpp,gu2017efficient,lihopp,guo2022clio}, as the implementation of memory disaggregation typically involves a concern about bandwidth and latency penalties over the network, which may adversely affect application performance~\cite{sysbalance}.

To date, previous research lacks a structured analytical method that can demonstrate which applications are performance constrained on the disaggregated memory system, how much the network performance penalty affects application performance, and what are essential metrics to assess the application performance impacts of a disaggregated memory system. This paper provides a structured system design model to explore the architecture design space, and its constraints. We provide several methods to visualize the design space and a methodology that could be adapted for a broader range of users, including vendors and application developers, to help to design new architectures or purchase future systems.

\section{System Architecture Design}\label{sec:sys}
\begin{figure}[!htb]
   \begin{minipage}{0.49\textwidth}
        \centering
        \includegraphics[width=0.9\textwidth]{ 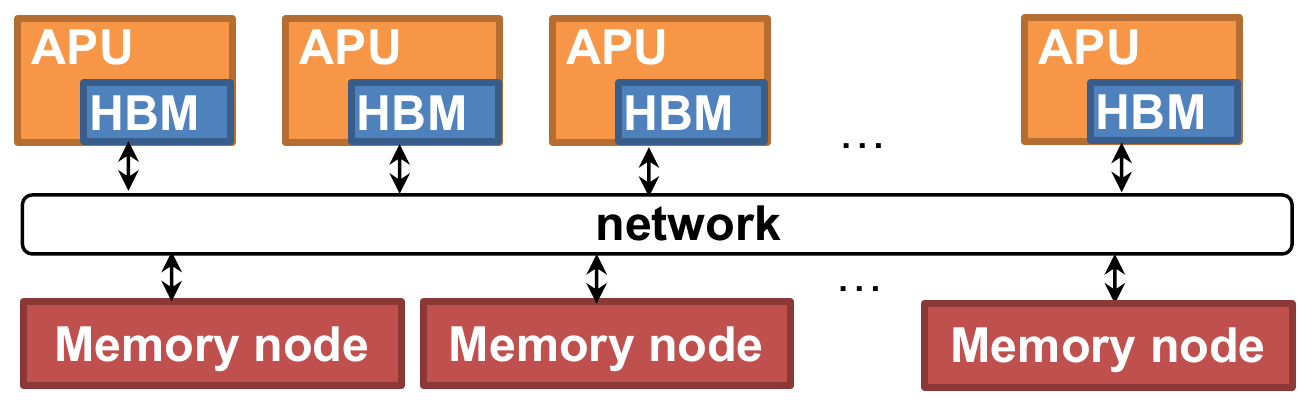}
        \caption{\label{fig:sys_abs} Conceptual disaggregated memory system architecture.}
   \end{minipage}
    \begin{minipage}{0.49\textwidth}
        \centering
        \includegraphics[width=0.9\textwidth]{ 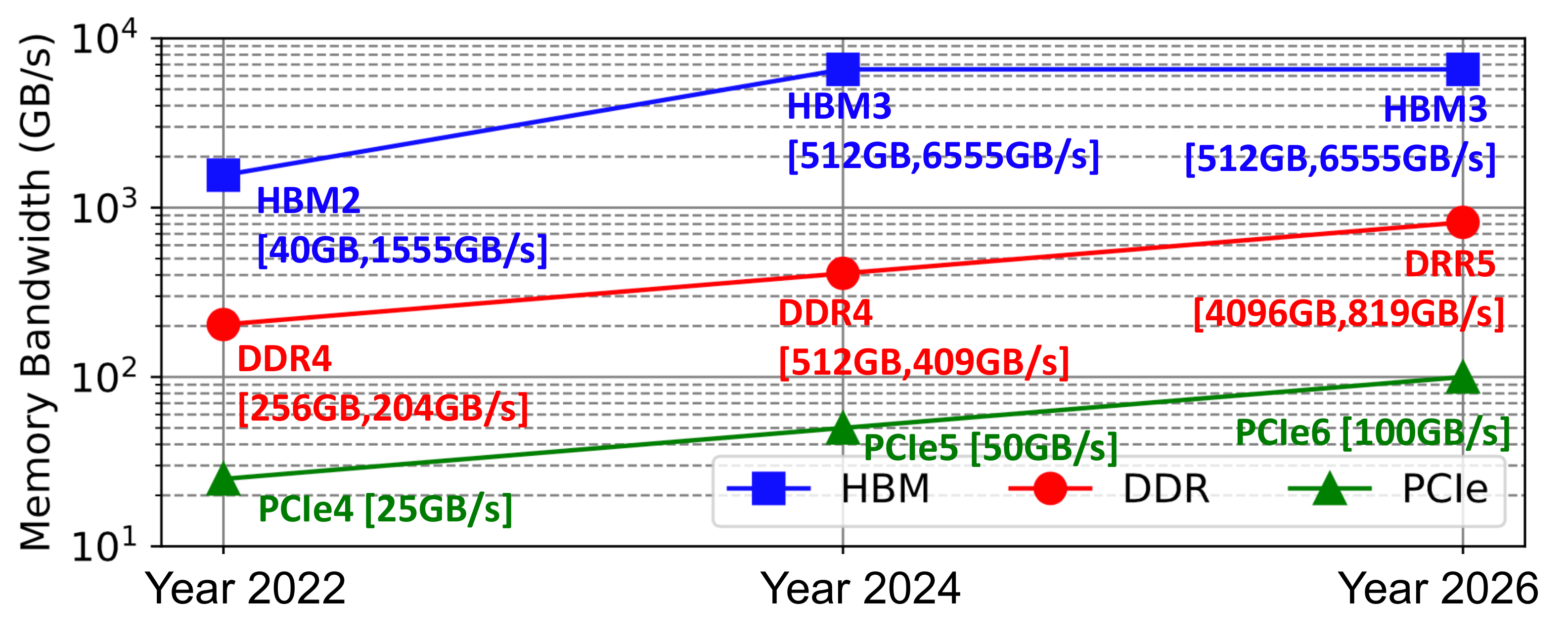}
        \caption{\label{fig:mem_tech_list} Memory bandwidth trends for HBM, DDR and PCIe between 2022 and 2026. Relative bandwidth improvements remain constant. The PCIe is the performance bottleneck for disaggregated systems.}
   \end{minipage}\hfill
\end{figure}

Understanding the emerging technology trends and capabilities in a future disaggregated memory system is necessary to assess the potential benefits and pitfalls. Fig.~\ref{fig:sys_abs} presents a basic network-attached disaggregated memory system schematic. We consider that such a system could have $C$ compute nodes and $M$ memory nodes. Each compute node consists of an accelerated processing unit (APU) with its own local high-bandwidth memory (HBM). An APU combines a CPU with a GPU onto a single silicon die, and both CPU and GPU share a common path to the remote memory. Future processor trends favor the APU because it addresses the bottleneck of data transfers between CPU and GPU ~\cite{daga2011efficacy,branover2012amd}. Each memory node is equipped with DDR memory as the remote memory. The compute nodes and memory nodes are connected via a network and is assumed to have one PCIe NIC. Thus, when data exceeds HBM capacity, data must be loaded from DDR via the network.

Fig.~\ref{fig:mem_tech_list} charts the memory bandwidth trends of HBM, DDR, and PCIe from today to the year 2026. Our assumptions for HBM3 include eight 16-Hi stacks, each with 64 GB capacity. We use the maximum capacity and bandwidth per DIMM (DDR4: 32 GB/DIMM and 25.6 GB/s/DIMM; DDR5: 256 GB/DIMM and 51.2 GB/s/DIMM) with a total of 16 DIMMs for DRAM memory. One can immediately notice that PCIe would eventually be the performance bottleneck on a disaggregated system since data needs to be loaded from DDR via the network.

In this section, we first propose a structured system architecture design methodology using the above assumptions to explore the potential and pitfalls of disaggregated memory. Secondly, we analyze the implication of bisection bandwidth, with two state-of-the-art network topologies -- Dragonfly and three-level fat tree, on disaggregated memory systems. 

\subsection{Available remote memory resources}
\begin{figure*}[!htb]
  	\begin{minipage}[t]{0.95\linewidth}
		\centering
        \subfloat[$\frac{C}{M}=\frac{1}{1}$: 100\% of one memory node capacity and 100\% of remote memory bandwidth]{
            \label{fig:threecases_1} 
            \includegraphics[width=0.32\textwidth]{ 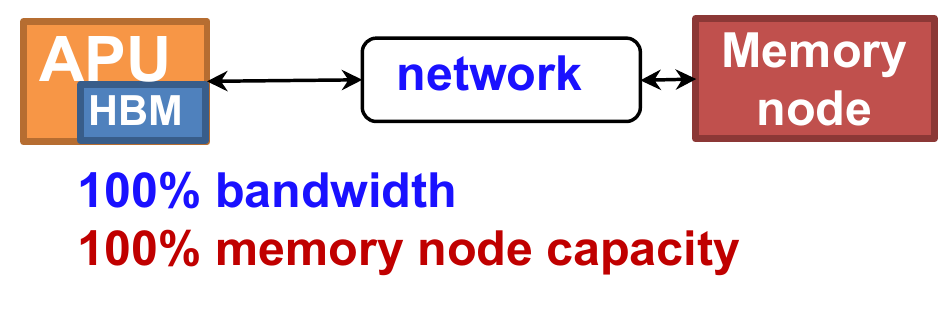}}
~ 
         \subfloat[$\frac{C}{M}=\frac{2}{1}$: 50\% of one memory node capacity and 50\% of remote memory bandwidth]{
            \label{fig:threecases_2} 
            \includegraphics[width=0.32\textwidth]{ 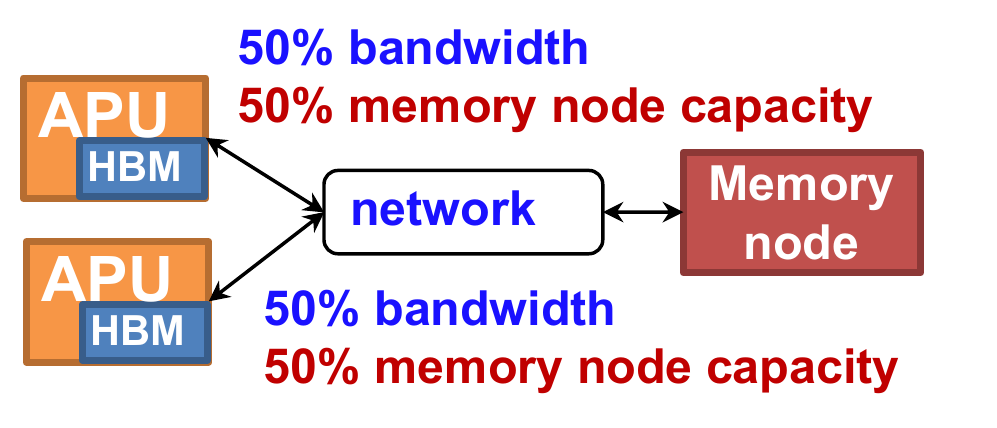}}
~ 
         \subfloat[$\frac{C}{M}=\frac{1}{2}$: 200\% of one memory node capacity and 100\% of remote memory bandwidth]{
            \label{fig:threecases_3} 
            \includegraphics[width=0.32\textwidth]{ 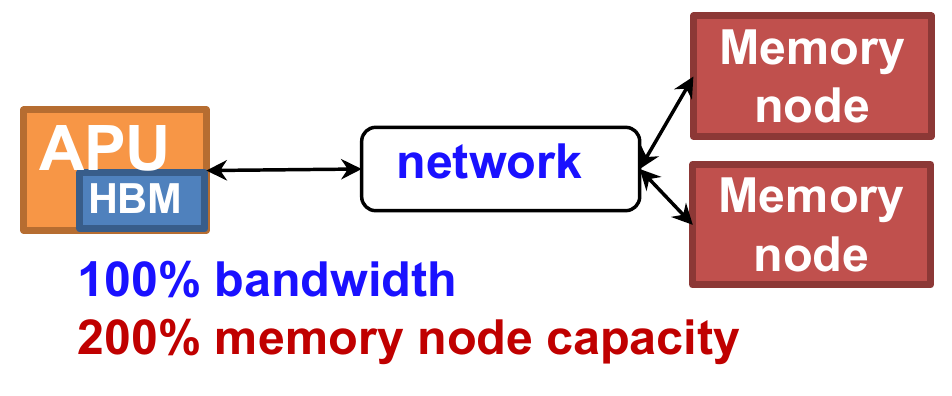}}

         \caption{\label{fig:threecases} Three use cases for network-attached disaggregated memory. Each varies the ratio of compute nodes (C) to memory nodes (M) and gives the available remote memory bandwidth (relative to one NIC) and remote memory capacity (relative to one memory node) available to each compute node. The flexibility of disaggregated memory allows systems to realize all three configurations simultaneously for different applications.}
	\end{minipage}
\end{figure*}

Fig.\ref{fig:threecases} visualizes three use cases that highlight the impact of system architecture on available remote memory capacity and memory bandwidth. Fig.~\ref{fig:threecases_1} highlights the simplest case ($\frac{C}{M}=\frac{1}{1}$), where each compute node is paired with one memory node to run the job. Each compute node would theoretically have access to the memory nodes full capacity and 100\% of the NIC's bandwidth as remote memory bandwidth. Unsurprisingly, in the case of $\frac{C}{M}=\frac{2}{1}$ in Fig.~\ref{fig:threecases_2}, each compute node has half the capacity and half the remote memory bandwidth. Interestingly, if $\frac{C}{M}=\frac{1}{2}$ as in Fig.~\ref{fig:threecases_3}, each compute node could access 200\% of a memory node's capacity but still only attain 100\% of the NIC bandwidth for remote memory bandwidth as bandwidth is constrained by the APU's NIC rather than the two memory node NICs. 

\begin{figure}[!tb]
  	\begin{minipage}[t]{0.98\linewidth}
		\centering
            \subfloat[Available remote memory capacity per compute node.]{
            \label{fig:heatmap_capacity} 
            \includegraphics[width=0.5\textwidth]{ 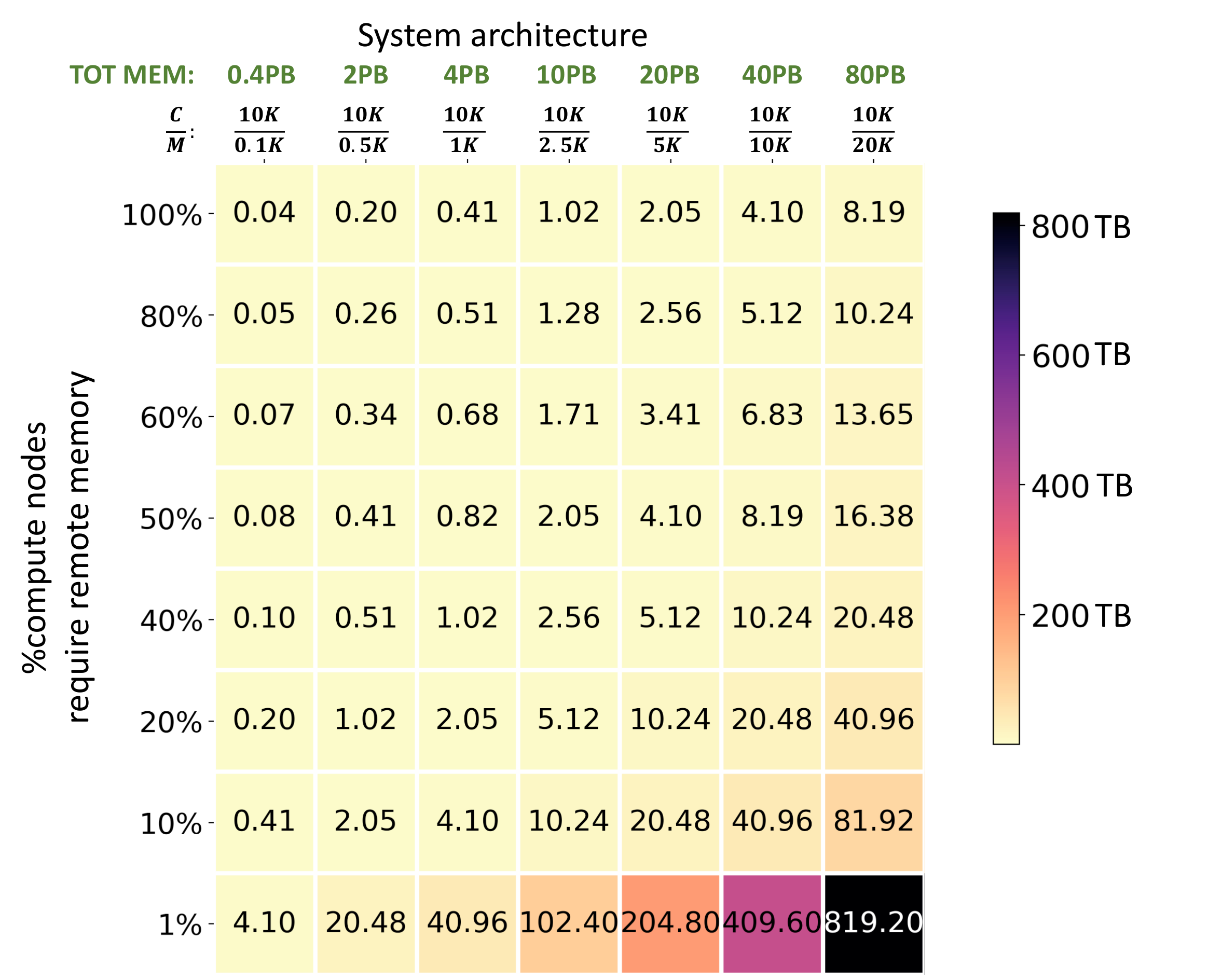}}
            \subfloat[Available remote memory bandwidth per compute node.]{
            \label{fig:heatmap_pcie6} 
            \includegraphics[width=0.5\textwidth]{ 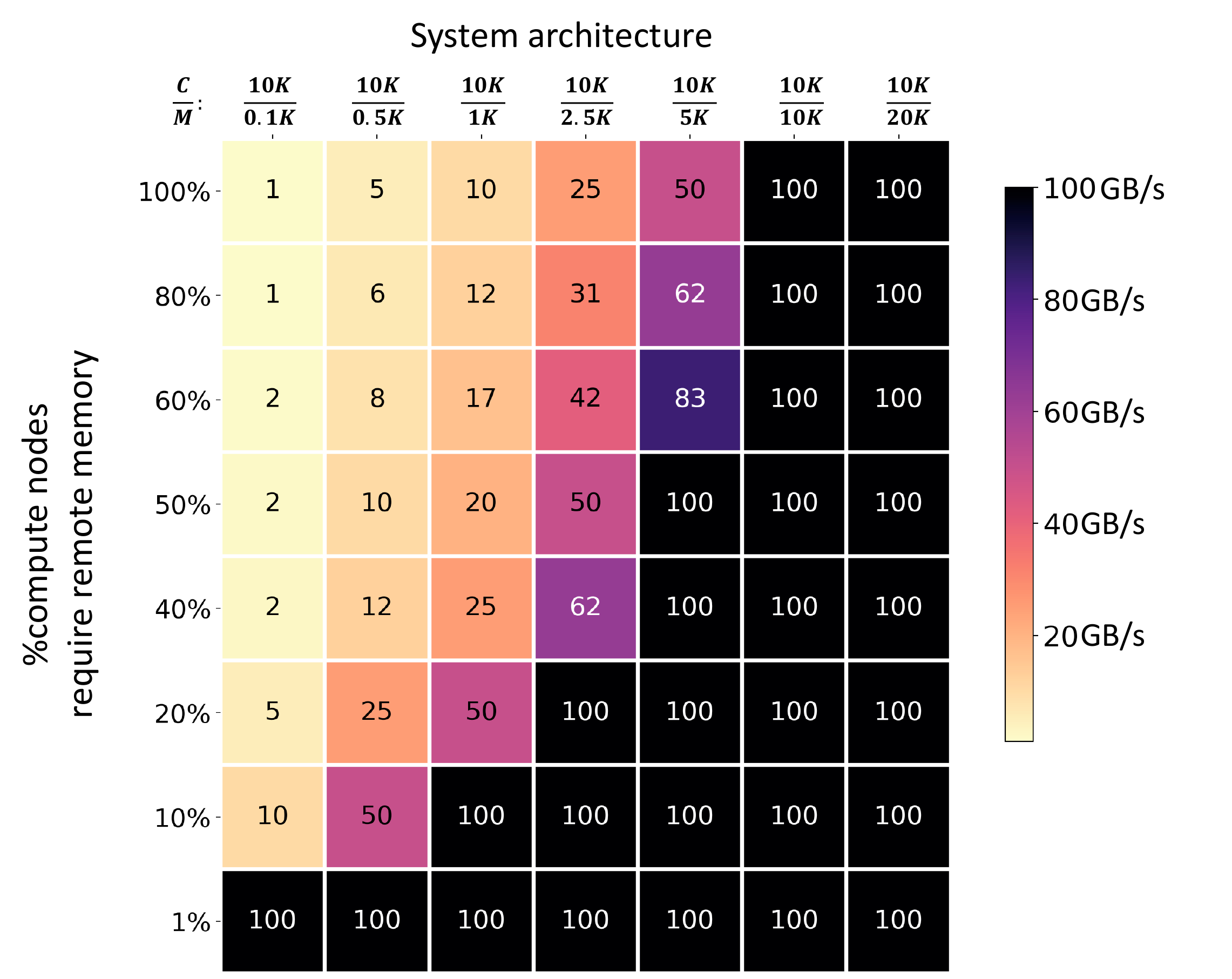}}

         \caption{\label{fig:heatmaps} Disaggregated memory system design space assuming fixed 10K (C) compute nodes and varying the number of (M) memory nodes, each with 4TB of DDR5 remote memory accessed through a PCIe6 NIC. Contention is reduced from left to right and from top to bottom. The y-axis shows the demand from compute nodes for remote memory and the x-axis shows the supply of memory nodes available. }
	\end{minipage}
\end{figure}

Following this method, we can build a design space with various ratios to describe the system constraints in terms of memory capacity and available remote memory bandwidth. We scale the building blocks to a modern-day HPC system and assume we have 10K compute nodes. The heat maps in Fig.~\ref{fig:heatmaps} present the (a) available remote memory capacity and (b) available remote memory bandwidth per compute node under different compute and memory node ratios, assuming one memory node capacity of 4TB. For the fixed number of compute nodes, Fig.~\ref{fig:heatmap_capacity} shows the available DDR5 remote memory capacity (TB) to the compute nodes with growing numbers of memory nodes (100 to 20K). The vertical axis is binned into the percentage of compute nodes that will require more resources than the local HBM memory and use the remote DDR memory, a value that will be specific to each HPC system and its workload. The available remote memory capacity per compute node becomes larger as we increase the number of memory nodes (moving left to right  of Fig.~\ref{fig:heatmap_capacity}). That is to say, there is less contention as we increase the number of memory nodes. Similarly, we can reduce contention as the number of compute nodes requiring remote memory decreases (moving top to bottom of Fig.~\ref{fig:heatmap_capacity}). For example, the first row represents the scenario where all the compute nodes require remote memory. If we focus on the sixth column where we have 10K DDR5 memory nodes ($\frac{C}{M}=\frac{1}{1}$), we see each compute node can access one memory node's capacity of 4TB. When decreasing the demand of the compute nodes for remote memory (moving down the column), at 50\% (5K) of the compute nodes requiring the remote memory, each compute node can then access 8TB of remote memory, which equals the capacity of two memory nodes. Correspondingly, Fig.~\ref{fig:heatmap_pcie6} presents the available remote memory bandwidth for the cases in Fig.~\ref{fig:heatmap_capacity}. Unlike memory capacity in Fig.~\ref{fig:heatmap_capacity}, memory bandwidth in Fig.~\ref{fig:heatmap_pcie6} will saturate at the compute node's peak NIC bandwidth regardless if one decreases the $\frac{C}{M}$ ratio (moving to the right) or one decreases the fraction of compute nodes requiring remote memory (moving down).

Determining an optimal system configuration relies on multiple factors specific to the HPC system workload (demand) and available budget (supply of memory nodes). Fig.~\ref{fig:mem_tech_list} suggests in the 2026 time frame, HBM3 could provide 0.5TB of local memory per node. Thus, in planning for the next machine, as a guiding principle, there should be enough memory nodes to provide more remote memory capacity per node than local memory capacity. As such, configurations in the upper left region of Fig.~\ref{fig:heatmaps} where memory node capacity is smaller than 0.5TB are wasteful architectures. 
Conversely, configurations on the right of the figure can be quite expensive as there are as many or more memory nodes than compute nodes (the network has 2-3$\times$ more endpoints).
Finally, although configurations in the bottom right provide 100s of TB per compute node, they can only access it at 100GB/s. As such, it will take minutes to hours to read all of remote memory once. Such architectural configurations may become impractical given the number of times an application might desire to read memory coupled with finite job run time limits.

\subsection{Bisection Bandwidth Implication}
The bisection bandwidth is the bandwidth available between two equal partitions of the network. It is used as an upper bound on collective communications, e.g., all-to-all. It also has been considered one of the key metrics that can affect applications' performance as HPC systems continue to increase in size~\cite{de2020depth,aroca2013bisection}. 
Such impact can be amplified on a disaggregated memory system because users can load data from anywhere in the entire memory pool rather than be confined to a single node. That being said, both inter-process communication (point-to-point and collective operations) and loading data from remote memory will contend for the PCIe NIC bandwidth. However, for cost and performance trade-offs, the bisection bandwidth of today's machines usually maintains only 28\% of the injection bandwidth. Thus, a critical decision in disaggregated memory system design is the interconnection resources, which determine the bisection bandwidth.

In this section, we analyze the bisection bandwidth of two state-of-the-art network topologies, the three-hop Dragonfly topology (Perlmutter~\cite{pm} and Frontier~\cite{frontier}) and the three-level Fat-tree topology (Summit~\cite{stunkel2020high}) to explore the bisection bandwidth. We then demonstrate the hardware limitation for intra-rack disaggregation and system-wide disaggregation. The intra-rack disaggregation~\cite{michelogiannakis2022case} (rack disaggregation) represents that applications can directly access all available memory inside its rack but would rely on RDMA for more memory beyond the rack. Similarly, system-wide disaggregation (global disaggregation) means that users can load data from anywhere across the entire system.

For a switch radix $k$, full-bandwidth three-level Fat-tree networks can scale to ${k^3}/{4}$ endpoints, and Dragonfly networks can scale to $k^4/64$ endpoints~\cite{stunkel2020high}. For our analysis, we consider using a 64-port switch since its scalability is sufficient for the expected number of endpoint links into the network on a 2026 disaggregated memory system for both topologies. We use a system size of 11,000 nodes (10,000 compute nodes and 1,000 memory nodes) as an example to predict the impact of bisection bandwidth. We assume that 10\% of the compute nodes will require remote memory for our machine configuration. Referencing the third column and seventh row of Fig~\ref{fig:heatmaps}, our machine configuration's maximum memory bandwidth per compute node is 100 GB/s.

Table~\ref{tab:bisection_config} presents the characteristics of the three-level Fat-tree and the three-hop Dragonfly using 64-port switches. According to the design discipline of the three-level Fat-tree on Summit, we can build a topology with 762 switches at the leaf level, of which forty-six ports per switch are used to connect the root level, and the rest of the sixteen ports are used to connect the endpoints. At the root level, we combine sixteen switches as one core switch and have sixteen core switches that are fully connected with each other. Thus, each core switch uses sixteen ports connecting to each other at the root level and uses the remaining forty-six ports connecting to the leaf-level switches. Ultimately, the three-level Fat-three topology requires 1018 switches ($762+16\times16$) with full bisection bandwidth. Note that the bisection of the three-level Fat-tree always achieves 100\% of the injection bandwidth.

\begin{table}[!htb]
\renewcommand{\arraystretch}{1.1}
\centering
\caption{Bisection bandwidth and its implication for a disaggregated system with 10K compute nodes and 1K memory nodes.} \label{tab:bisection_config}
\resizebox{0.95\textwidth}{!}{
\begin{tabular}{lll|ll|llll}
\toprule[1.2pt] 
\hline
\textbf{Machine} & \textbf{Topology} & \textbf{Config} & \multicolumn{2}{c|}{\textbf{intra-group/rack disaggregation} }  & \multicolumn{4}{c}{\textbf{inter-group/global disaggregation} }  \\ 
  &  &  & \textbf{bisection} & \textbf{taper}~\label{tab:col:intraimpl} & \textbf{bisection} &\textbf{taper}~\label{tab:col:interimpl} & \textbf{\#switches} & \textbf{\#Total links} \\ \hline
Perlmutter & Dragonfly & 24 groups x 16switches & 25GB/s & 100\% of PCIe4 & 7GB/s & 28\% of PCIe4 (6 links/pair) & 384 & 3312 \\ \hline
\multirow{4}{*}{Disaggregated} & \multirow{4}{*}{Dragonfly} & \multirow{4}{*}{24 groups x 32 switches} & \multirow{4}{*}{100 GB/s} &   & 9 GB/s & 9\% of PCIe6 (4 links/pair) & 768 & 2208 \\
&  &  &  &100\% of PCIe6 & 28 GB/s & 28\% of PCIe6 (12 links/pair) & 768 & 6624 \\ 
&  &  &  & 1 link/pair& 50 GB/s & 50\% of PCIe6 (21 links/pair) & 768 & 11592 \\
&  &  & & & 100 GB/s & 100\% of PCIe6 (43 links/pair) & 768 & 23736 \\ 
 \hline
 \multirow{3}{*}{Disaggregated} & \multirow{3}{*}{Dragonfly} & \multirow{3}{*}{48 groups x 16 switches} & \multirow{3}{*}{50 GB/s} &  & 28 GB/s & 28\% of PCIe6 (3 links/pair) & 768 & 6768 \\ 
  &  &  & & 50\% of PCIe6 & 56 GB/s & 56 \% of PCIe6 (6 links/pair) & 768 & 13536 \\ 
  &  &  & & 1 link/pair & 100 GB/s & 100\% of PCIe6 (11 links/pair) & 768 & 24816 \\ \hline
 Disaggregated & Fat-tree & three-level & 100GB/s & 100\% of PCIe6 & 100GB/s & 100\% of PCIe6 & 1018 & 11776 (\#links between levels) \\ 
 \hline
\bottomrule[1.4pt]
\end{tabular}}
\end{table}

For the three-hop Dragonfly, we can either scale the number of groups or switches per group to build a larger dragonfly network.
Throughout the paper, we refer to the intra-group bisection bandwidth as the available remote memory bisection bandwidth for rack disaggregation, and inter-group bisection bandwidth as the  available remote memory bisection bandwidth for global disaggregation. 
Specific to our machine configuration, we consider two choices: 24 groups with 32 switches each and 48 groups with 16 switches each. 
Even though the two settings have the same number of switches, they have very different intra- and inter-group bisection bandwidths and incur different costs. 
For example, in the 24-group setting,
one can only achieve 9\% of the injection bandwidth if it keeps a similar cost to Perlmutter. 
Another choice is to triple the number of inter-group links to maintain the 28\% bisection bandwidth tapering as Perlmutter~\cite{perlmutter}.

Undoubtedly, one can achieve high bisection bandwidth with more link costs, i.e., 100\% of the PCIe6 NIC, but the cost is extremely high --- four times higher than the 28\% configuration. The 48-group setting has a similar cost for 28\% tapering, while it can only achieve 50\% intra-group bisection bandwidth if it is limited to one link per intra-group pair.
The implication of bisection bandwidth can be immediately noticed: the bisection network will reduce the available remote memory bandwidth, and the degree of reduction varies with different configurations.

\begin{figure}[!tb]
  	\begin{minipage}[t]{0.98\linewidth}
		\centering
            \subfloat[Injection bandwidth]{
            \label{fig:heatmap_bisection_injection} 
            \includegraphics[width=0.32\textwidth]{ 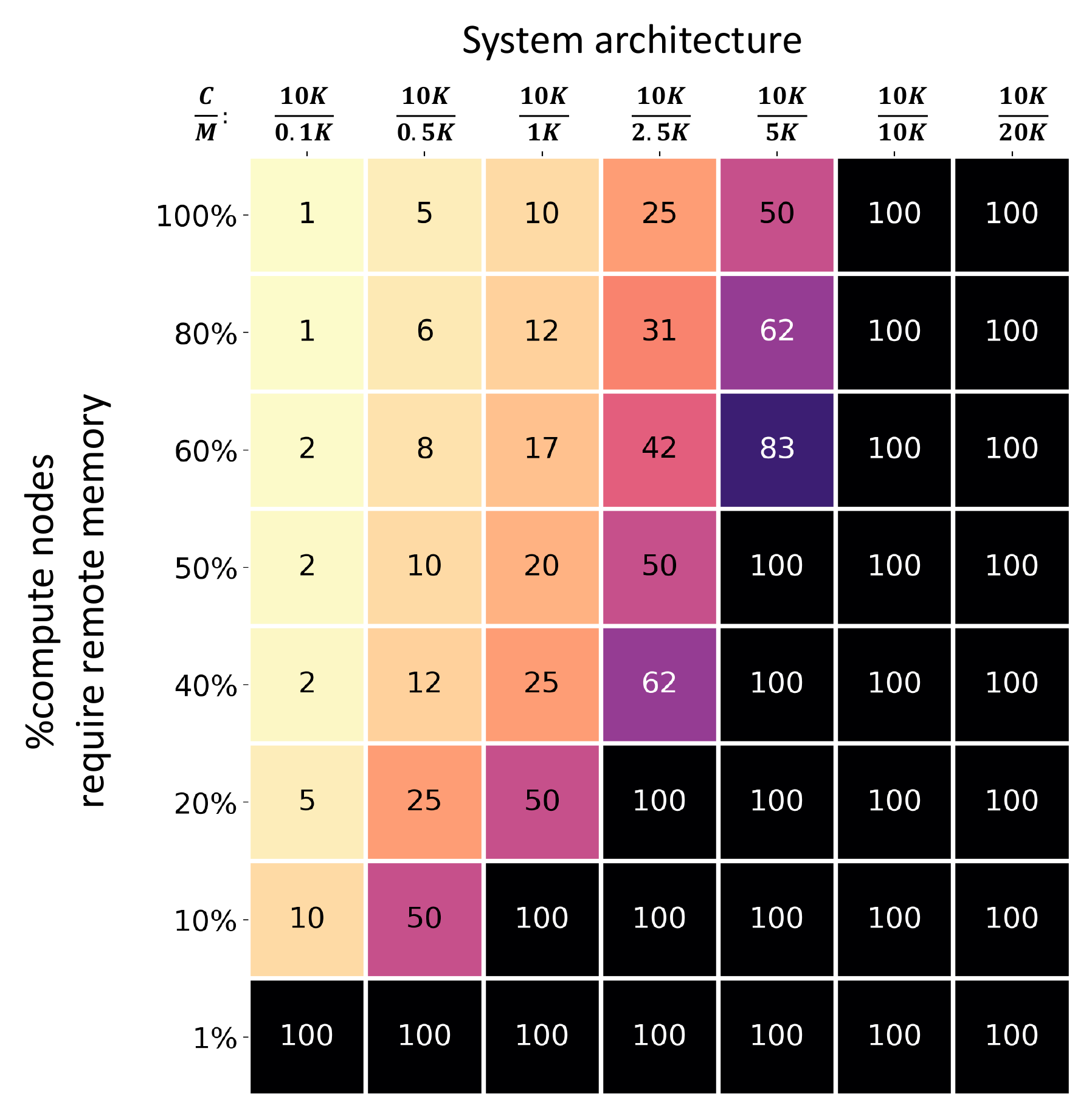}}
            \subfloat[Rack bisection]{
            \label{fig:heatmap_bisection_5injection} 
            \includegraphics[width=0.32\textwidth]{ 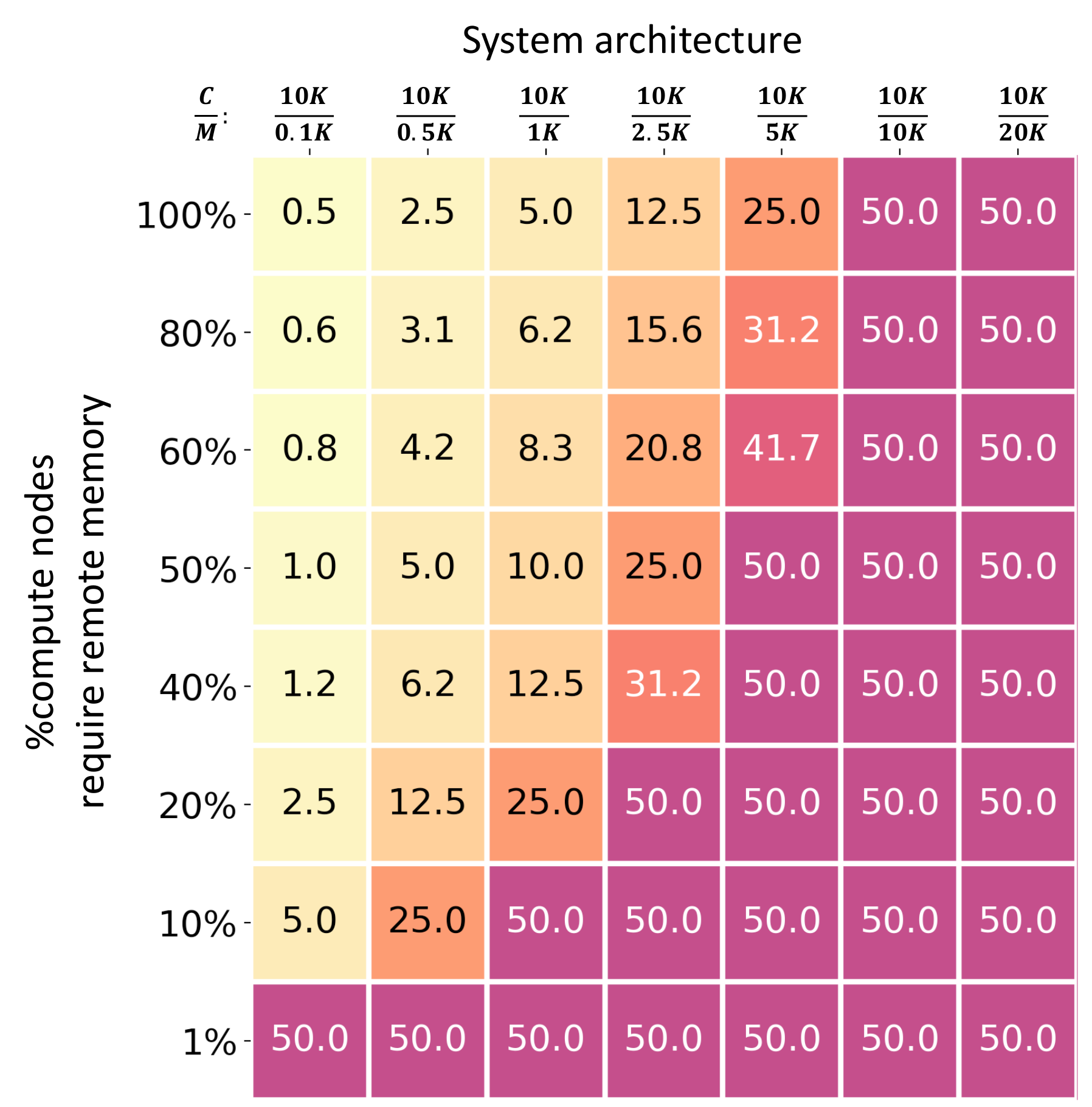}}
            \subfloat[Global bisection]{
            \label{fig:heatmap_bisection_9injection} 
            \includegraphics[width=0.4\textwidth]{ 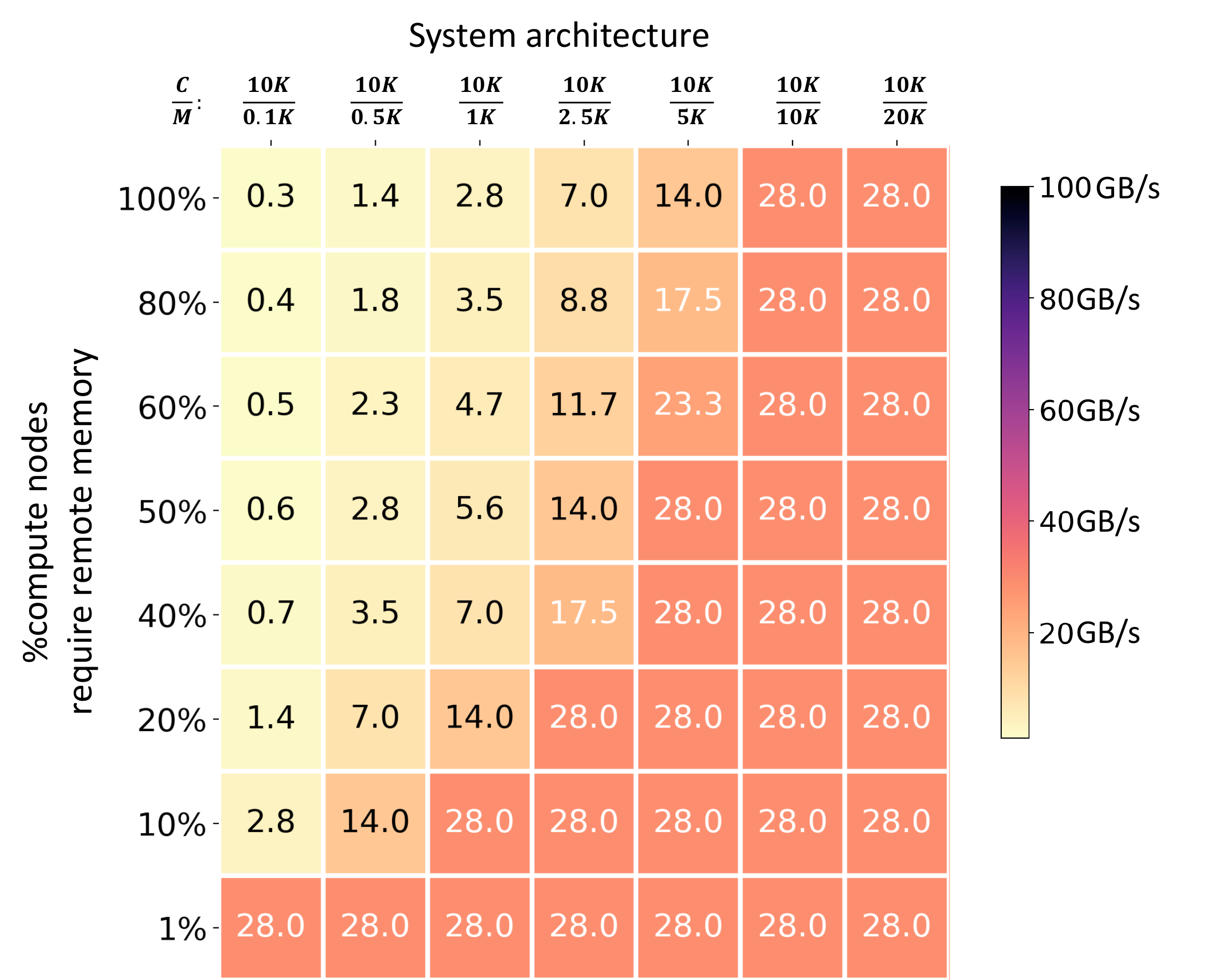}}

         \caption{\label{fig:heatmaps_bisection} Bisection bandwidth implications on disaggregated memory system design space assuming the bisection bandwidth is 100\%, 50\% and 28\% of the injection bandwidth. An intra-rack disaggregated memory system will see its available remote memory bandwidth halved by the bisection network in Fig~\ref{fig:heatmap_bisection_5injection}. A global disaggregated memory system's  available remote memory bandwidth is even lower in Fig.\ref{fig:heatmap_bisection_9injection}. Bigger system sizes result in more expensive networks: more switches and links to maintain 28\% taper.}
	\end{minipage}
\end{figure}

Figure~\ref{fig:heatmaps_bisection} highlights the bisection bandwidth implications on the disaggregated memory system design space, assuming the bisection bandwidth is 100\%, 50\% and 28\% of the injection bandwidth. Note that each system size (x-axis) may require a different network configuration. A bigger system size will need a more expensive network, e.g., a system with 30K nodes (C:M=10K:20K, 74~groups$\times$32~switches~=~2368 switches) needs 3$\times$ higher cost than 11K nodes (C:M=10K:1K, 768 switches), and 165$\times$ more links to maintain 28\% taper. One would imagine an intra-rack disaggregated memory system will see its available remote memory bandwidth halved by the bisection network (50\% taper) in Fig~\ref{fig:heatmap_bisection_5injection}. Fig.\ref{fig:heatmap_bisection_9injection} shows that a globally disaggregated memory system's available remote memory bandwidth is less than  28\% of its injection bandwidth.

Like the system architecture design, determining the network configuration relies on multiple factors specific to the HPC system workload (demand) and available budget (supply of switches and links). The guiding principle is that there should be enough remote memory bandwidth to support collective operations through the network with a minimal negative impact on its workloads. 

\section{Application Characterization}\label{sec:appcharacter}
In this section, we propose a memory Roofline model to evaluate and visualize the performance bottlenecks of applications running on a disaggregated memory system.
Disaggregated memory promises to improve system-wide memory utilization, but individual application performance is of equal concern. Prior work argues that disaggregation comes with substantial bandwidth and latency penalties to applications~\cite{peng2020memory}. However, such conclusions are derived assuming current technologies and lack consideration of emerging technologies in the future. To analyze the impact on individual applications in the near future, we introduce the local-to-remote memory access ratio (L:R) metric to characterize application performance on a disaggregated memory system. We then correlate the metrics using a memory Roofline plot to provide a generalized framework to evaluate and visualize the performance bottlenecks of applications running on a disaggregated memory system.

The traditional Roofline model~\cite{williams2009Roofline} characterizes an application's performance (GFLOP/s) as a function of its arithmetic intensity (FLOPs executed per Byte moved). It provides a quick visual comparison of the application performance compared against the bounds set by the peak compute performance (GFLOP/s) and the peak memory bandwidth of the target architecture (GB/s) to determine what is limiting performance: memory or compute.
Following the methodology of the traditional Roofline model, our new memory Roofline model characterizes an application's sustained memory performance (GB/s) as a function of its local and remote memory access ratio (L:R), the peak local memory bandwidth, and the peak remote memory bandwidth.
An application's L:R on a disaggregated memory system could be considered as the ratio of HBM data movement (local) to the DDR data movement (remote over PCIe) or even the HBM to file size ratio when examining applications using memory nodes as a private file system.
Applications with an L:R data movement ratio greater than the system's local:remote bandwidth ratio can effectively hide the slow remote (disaggregated) memory bandwidth behind a multitude of fast, local memory accesses.

\begin{figure}[!tb]
  	\begin{minipage}[t]{0.98\linewidth}
		\centering
            \subfloat[Available remote memory injection bandwidth.]{
            \label{fig:memoryRoofline} 
            \includegraphics[width=0.48\textwidth]{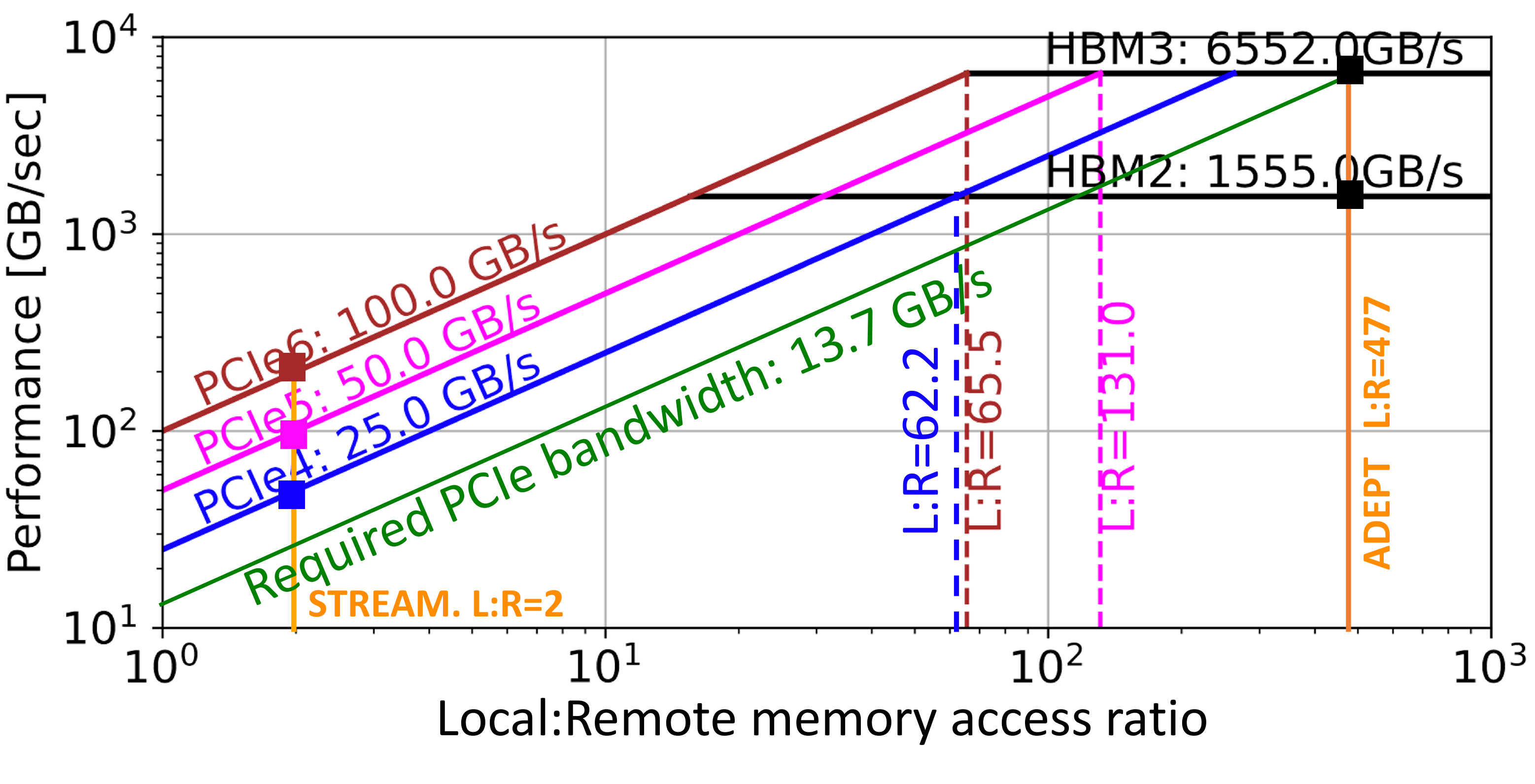}}
            \subfloat[Available remote memory bisection bandwidth.]{
            \label{fig:memoryRoofline_bisection} 
            \includegraphics[width=0.48\textwidth]{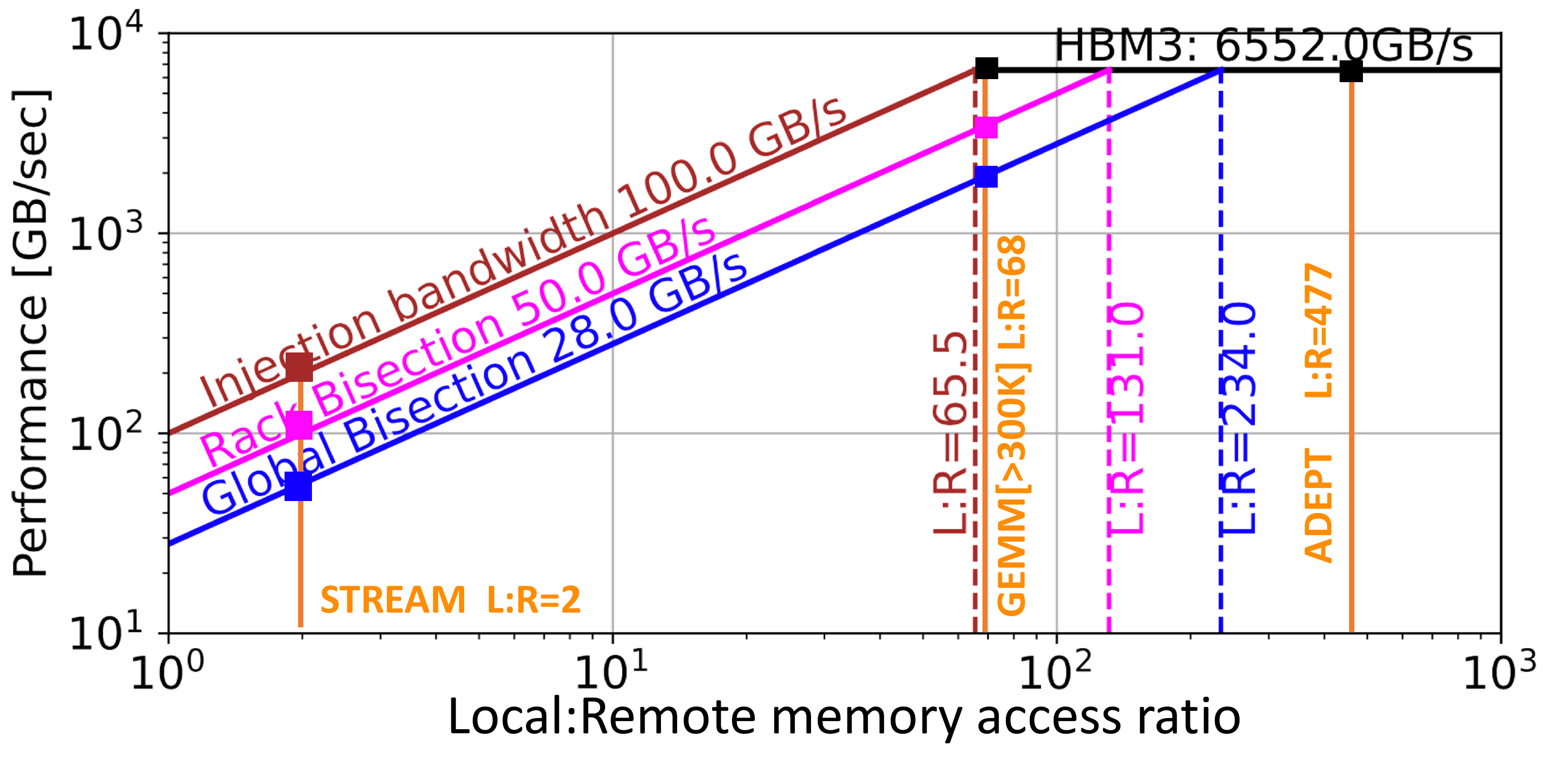}}

         \caption{\label{fig:memoryrooflineall}  Memory Roofline model characterizing an application's memory access performance (GB/s) as a function of its local to remote memory access ratio (L:R). A high L:R ratio is critical in mitigating the performance penalties of disaggregated memory. Observe ADEPT is insensitive to full bandwidth disaggregated memory while STREAM is penalized by it. GEMM is insensitive to the injection bandwidth but is penalized by bisection bandwidth. } 
	\end{minipage}
\end{figure}

Fig.~\ref{fig:memoryRoofline} presents the memory Roofline model using future HBM (local) and PCIe (remote) bandwidths.
One quickly observes the visual similarity to the traditional Roofline model with local bandwidth replacing the traditional peak GFLOP/s plateau and remote bandwidths replacing the traditional memory diagonals.
We observe an HBM3:PCIe6 machine balance of 65.5 --- the ratio of data movement that results in equal time for local and remote transfers. This ratio is very close to today's HBM2:PCIe4 machine balance of 62.2. This suggests future hardware trends will not detract from the efficacy of disaggregated memory.

Applications like ADEPT with an L:R ratio of nearly 500 (far greater than 65.5) are insensitive to memory disaggregation, dominated by on-node performance, and will use less than 14\% of the available PCIe bandwidth (green diagonal line).
Conversely, applications like STREAM with a theoretical L:R ratio of 2 will see their performance limited and degraded by disaggregated memory bandwidth.

Figure~\ref{fig:memoryRoofline_bisection} shows performance as a function of the  bisection bandwidth tapering relative to injection bandwidth.  
Bisection bandwidth shifts the machine balance to the right, e.g., a  50\% tapering increases the machine balance from 65.5 to 131 local words per remote word. One could imagine building an intra-rack disaggregated memory system with 50\% tapering (pink in Figure~\ref{fig:memoryRoofline_bisection}) and a global disaggregated system with 28\% tapering (blue in Figure~\ref{fig:memoryRoofline_bisection}). 
A hypothetical GEMM with matrix dimension of about 300K $\times$ 300K will be limited by bisection bandwidth and unable to even use the full local HBM bandwidth. Conversely, applications like ADEPT with a high L:R are insensitive to reasonable rack- and global bisection bandwidths. 
Ultimately, the applications with L:R ratio less than 131 will see rack bisection bandwidth as a larger bottleneck that local memory bandwidth, while applications with an L:R ratio less than 234 running on a globally disaggregated memory will see global bisection bandwidth as bigger performance impediment than local bandwidth.

Ultimately, increases in network bandwidth shift the machine balance to the left (decreasing the number of applications penalized by disaggregation), while increases in HBM bandwidth shift the machine balance to the right (increasing the number of applications penalized by disaggregation).
Whereas the latter simply scales the cost of each node, increasing bisection bandwidth can scale superlinearly with the number of nodes. As such, shifting the local:global balance to the left can be cost prohibitive for large HPC and cloud systems.

\section{Application Case Studies}\label{sec3}
In this section, using our methodology, we evaluate the efficacy of our proposed disaggregated memory system on a variety of application workloads.

\subsection{Disaggregated System Configurations}
Recall the HPC system described in Section~\ref{sec:sys} to select a machine configuration. A with 10,000 compute nodes with 512 GB of HBM3 local memory capacity, accessing DDR5 remote memory nodes via PCIe6-connected NICs. As previous studies showed only 15\% of the workloads use 75\% of the node memory~\cite{austin2020}, we conservatively assume that at any instant, 10\% of the compute nodes will require remote memory for our machine configuration. Referencing Fig~\ref{fig:heatmap_capacity}, at 10\%, we could choose 500 memory nodes or more with DDR5 memory (x-axis) to ensure each compute node has access to remote memory greater than the local HMB3 memory. Including the memory bandwidth information from Fig~\ref{fig:heatmap_pcie6}, the maximum memory bandwidth per compute node peaks at 1000 memory nodes. Purchasing more memory nodes would only add additional capacity and cost, not additional memory bandwidth. For the configuration of 10,000 compute nodes accessing an aggregate four petabytes of DDR5 memory on 1000 memory nodes, we see from Fig.\ref{fig:heatmaps} that each of the compute nodes requiring remote memory can access, on average, four terabytes of remote memory with a peak remote memory bandwidth of 100 GB/s. 

\subsection{Application Characteristics}
The examined case studies needed various approaches to measure or estimate the local and remote memory accesses (L:R) due to the diversity of their applications as Table~\ref{tab:methods} presents. This section summarizes the high-level methods for calculating L:R for each workload. It's important to note that throughout the paper, we assume that each application will maintain its current conceptual approach to leveraging data locality and expressing data movement even if the mechanisms are expressed differently in a disaggregated memory architecture.

\begin{table*}[!htbp]
\centering
\caption{Approaches for characterizing the local and remote memory accesses (L:R) and remote memory capacities} \label{tab:methods}
\begin{minipage}{\textwidth}
\resizebox{\textwidth}{!}{
\centering
\begin{tabular}{l|l|l|l|l}
\toprule[1.2pt] 
\hline
Workloads& Applications & Local memory accesses  & Remote memory accesses & Remote memory capacity \\ \hline
\multirow{3}{*}{AI training} & ResNet-50 & \multirow{3}{*}{memory usage of the activation layer}  & \multirow{3}{*}{sample batch size} & \multirow{3}{*}{Training set size} \\ 
&DeepCAM &  &  & \\ 
&CosmoFlow &  &  & \\ 
\hline
\multirow{2}{*}{Data analysis} & DASSA & analytical modeling &\multirow{2}{*}{analytical modeling} & \multirow{2}{*}{input size}\\  \cline{2-3}
& TOAST & Intel VTune &  & \\  \hline
\multirow{2}{*}{Genomics} & ADEPT (w/wo traceback) & analytical modeling & \multirow{2}{*}{analytical modeling} & \multirow{3}{*}{matrix  size} \\ \cline{2-3}
& EXTENSION & NVIDIA NSight compute &  & \\\cline{1-4}
Protein & PASTIS & NVIDIA NSight compute & analytical modeling & \\  \hline
FUSION  & SuperLU & analytical modeling & analytical modeling & analytical modeling \\ \hline
MFDn  & Eigensolver & analytical modeling & analytical modeling & analytical modeling \\ \hline
\multirow{2}{*}{Traditional HPC } & GEMM & \multirow{2}{*}{analytical modeling} & \multirow{2}{*}{analytical modeling} & \multirow{2}{*}{analytical modeling}\\ \cline{2-2}
& STREAM &  &  & \\ \hline
\multicolumn{2}{l|}{NVIDIA NSight compute Metrics} & \multicolumn{3}{l}{ 32B*(dram\_\_sectors\_read.sum + dram\_\_sectors\_write.sum)} \\ \hline
\multicolumn{2}{l|}{Intel VTune Metrics } & \multicolumn{3}{l}{ 64B*sum(UNC\_M\_CAS\_COUNT.WR[UNIT0-7],UNC\_M\_CAS\_COUNT.RD[UNIT0-7])}\\
\hline
\bottomrule[1.4pt]
\end{tabular}
}
\end{minipage}
\label{tab:app_character}
\end{table*}

\textbf{Artificial intelligence (AI) training workloads.~}
AI is an area of increasing scientific interest with growing computational demands ~\cite{austin2020} and drives future DOE investments in HPC platforms~\cite{baker2019workshop}. We focus on training workloads, which are more computationally expensive and require a larger memory capacity than inference. We demonstrate the benefit of a disaggregated memory system using three AI training workloads: CosmoFlow~\cite{mathuriya2018cosmoflow} and
DeepCAM~\cite{kurth2018exascale} from the MLPerf HPC benchmark suite~\cite{mlperftraining}, and
a well-established image classification model, ResNet-50~\cite{he2016deep} from the MLPerf Training benchmark suite~\cite{mattson2020mlperf}. The actual computation and memory characteristics of the three AI training workloads come from Ibrahim~\etal~~\cite{ibrahim2021architectural} and listed in Table~\ref{tab:ai_training_apps}. The local:remote memory ratio is calculate by dividing the measured FLOP:Sample Byte by the measured FLOP:HBM Byte. All the numbers reported in Table~\ref{tab:ai_training_apps} refer to the memory per job.

\begin{table}[htb]
\renewcommand{\arraystretch}{1.1}
\centering
\caption{Computation and Memory Characteristics}
\resizebox{0.48\textwidth}{!}{
\begin{tabular}{llll}
\toprule[1.2pt] 
\hline
& ResNet-50 & DeepCAM & CosmoFlow   \\ \hline
Training set size~\cite{ibrahim2021architectural} & 0.15 TB & 8.8 TB & 5.1 TB  \\ \hline
FLOP:HBM Byte~\cite{ibrahim2021architectural} & 55.35 & 55.5 & 38.6 \\ \hline
FLOP:sample Byte~\cite{ibrahim2021architectural} & 221,000 & 107,000 & 15,400   \\ \hline
Local:Remote memory access ratio & 3993 & 1927 & 399 \\ \hline
\bottomrule[1.4pt]
\end{tabular}}
\label{tab:ai_training_apps}
\end{table}

\textbf{Data analysis workloads.~}
Data analysis applications are a growing workload in HPC facilities~\cite{austin2020}. To showcase disaggregated memory benefits, we use two data analysis software frameworks, DASSA~\cite{dong2020dassa} and TOAST~\cite{toast}. DASSA~\cite{dong2020dassa} is a distributed acoustic sensing (DAS) data storage and analysis framework for geophysicists to perform DAS data analysis on HPC systems. We use a real DAS data analysis case for earthquake detection via local similarity. We use analytical modeling to estimate the L:R and refer its input file size as the remote memory capacity requirement.

TOAST~\cite{toast} is a software framework designed for simulation and data reduction from telescope receivers that acquire time streams of individual detector responses. Here we use a satellite telescope benchmark as an example to show the implication of memory disaggregation. The core computation in the satellite telescope benchmark is the PCG solver. We profile its DRAM data movement using Intel VTune on one Cori Haswell~\cite{cori} node as its local memory accesses and refer its input file size as the remote memory capacity requirement.

\textbf{Genomics workloads.~}
With the rapid development of genome sequencing technologies, it is now possible to sample and study genomes at an unprecedented scale. MetaHipMer~\cite{hofmeyr2020terabase} is a large-scale
metagenome assembler that can leverage the large memory and
compute capacities of supercomputers to co-assemble terabase-scale datasets. We use three important kernels in MetaHipMer, ADEPT~\cite{adept_awan} with and without traceback and EXTENSION~\cite{awan2021accelerating} to understand their potential on a disaggregated memory system. We use analytical modeling to calculate the L:R of ADEPT with and without traceback kernels. We use NVIDIA NSight compute~\cite{nsight} to collect the HBM data movement for single extension on Cori GPU~\cite{corigpu}, and then  multiply that with 45 million extensions as its local memory access. We use analytical modeling to estimate the remote memory capacities for all three kernels.

\textbf{Protein similarity search workloads.~}
Bioinformatics applications have increasingly turned to HPC solutions for solving big problems with reasonable time-to-solution.
Especially in metagenomics research, the scale of the data often requires memory and compute resources that are beyond what serial systems can provide.
An important task that forms the backbone of many bioinformatics workflows is the alignment of a set of given sequences against a reference database.
PASTIS~\cite{Selvitopi2020} is a distributed-memory many-against-many search tool specifically developed for protein sequences.
This search requires a lot of memory and its memory complexity grows quadratically with the number of sequences while being compute-intensive.
For batch pairwise alignments required by the protein similarity search, PASTIS uses SeqAn~\cite{Doring2008} for CPUs and ADEPT~\cite{adept_awan} for GPUs. We use NVIDIA NSight compute~\cite{nsight} to collect the HBM data movement as the local memory access and use analytical modeling to estimate the remote memory capacities.

\textbf{FUSION workload.} SuperLU\_DIST is a distributed memory sparse direct solver for large sets of linear equations. It is used as a preconditioner within an iterative solver in fusion simulation applications~\cite{Jardin2008M3dc1,sovinec04}.
In practice, one factors the system (SpLU) and performs a pair of triangular solves (SpTS) for each of the 100+ iterations of the iterative solver.
These two components dominate the run time. To amortize SpLU time, one can factor the system once and then use those factors (as a preconditioner) over the course of multiple time steps (the system didn't change much).
We use analytical modeling~\cite{grigori2010brief} to estimate the L:R, and refer to the factored matrix size as its memory requirement. 

\textbf{MFDn workload.} The LOBPCG eigensolver dominates the run time of the 2-body forces Many-body Fermion Dynamics for nuclear (MFDn) application~\cite{maris2022accelerating}. MFDn is application used to simulate the properties of atomic nuclei. LOBPCG performs a sparse matrix–matrix multiplication (SpMM) with a varied number of right-hand sides. We use analytical modeling to estimate the L:R, and refer to half the input matrix size as its memory requirement because the input matrix is symmetric.

\textbf{Traditional HPC Workload Bookends.~}
Traditional HPC workloads are designed for distributed-memory systems. They can sometimes scale to thousands of and even millions of cores~\cite{ding2021message,ding2020leveraging,fu2017redesigning}. They can distribute the memory footprint, and can fit in the 512 GB HBM3 in a 2026 disaggregated system which is larger than the currently provided node-local DDR (256 GB DDR on a 2021 HPC system). We use GEMM~\cite{gemm} and STREAM~\cite{stream} as two representative benchmarks to show the implications as the data size grows. We use analytical modeling to estimate the L:R for GEMM and STREAM. Note that STREAM can be a proxy for giant AI=O(1) linear solvers (stencil/sparse) without any multiphysics/AMR.

\subsection{Application Analysis}
\begin{figure}[htb]
    \begin{minipage}[t]{0.98\linewidth}
		\centering
            \subfloat[Rack disaggregation. Each rack contains 20 memory nodes (48 dragonfly groups).]{
            \label{fig:app_metric_rack} 
            \includegraphics[width=0.48\textwidth]{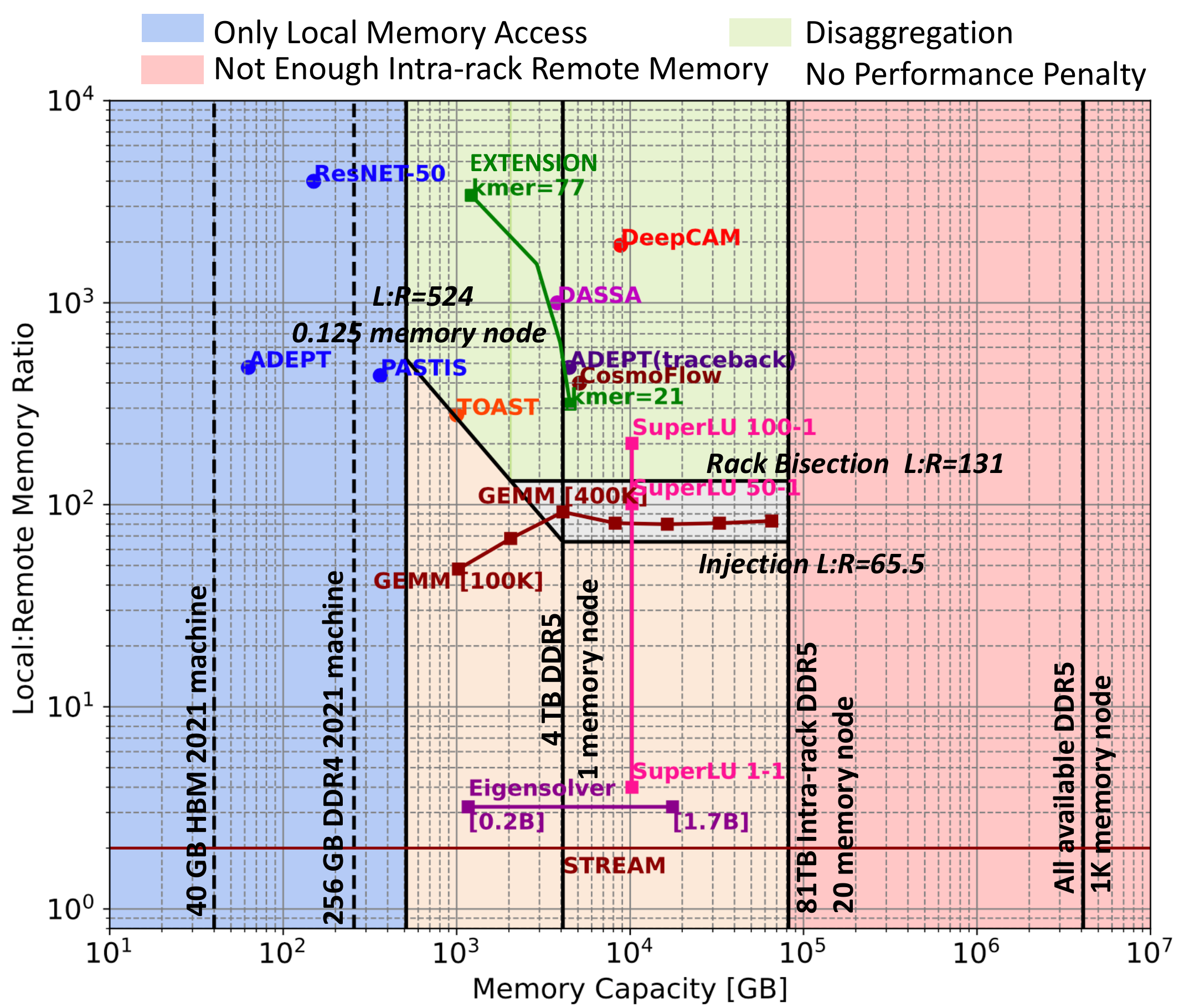}}
            \subfloat[Global disaggregation]{
            \label{fig:app_metric_global} 
            \includegraphics[width=0.48\textwidth]{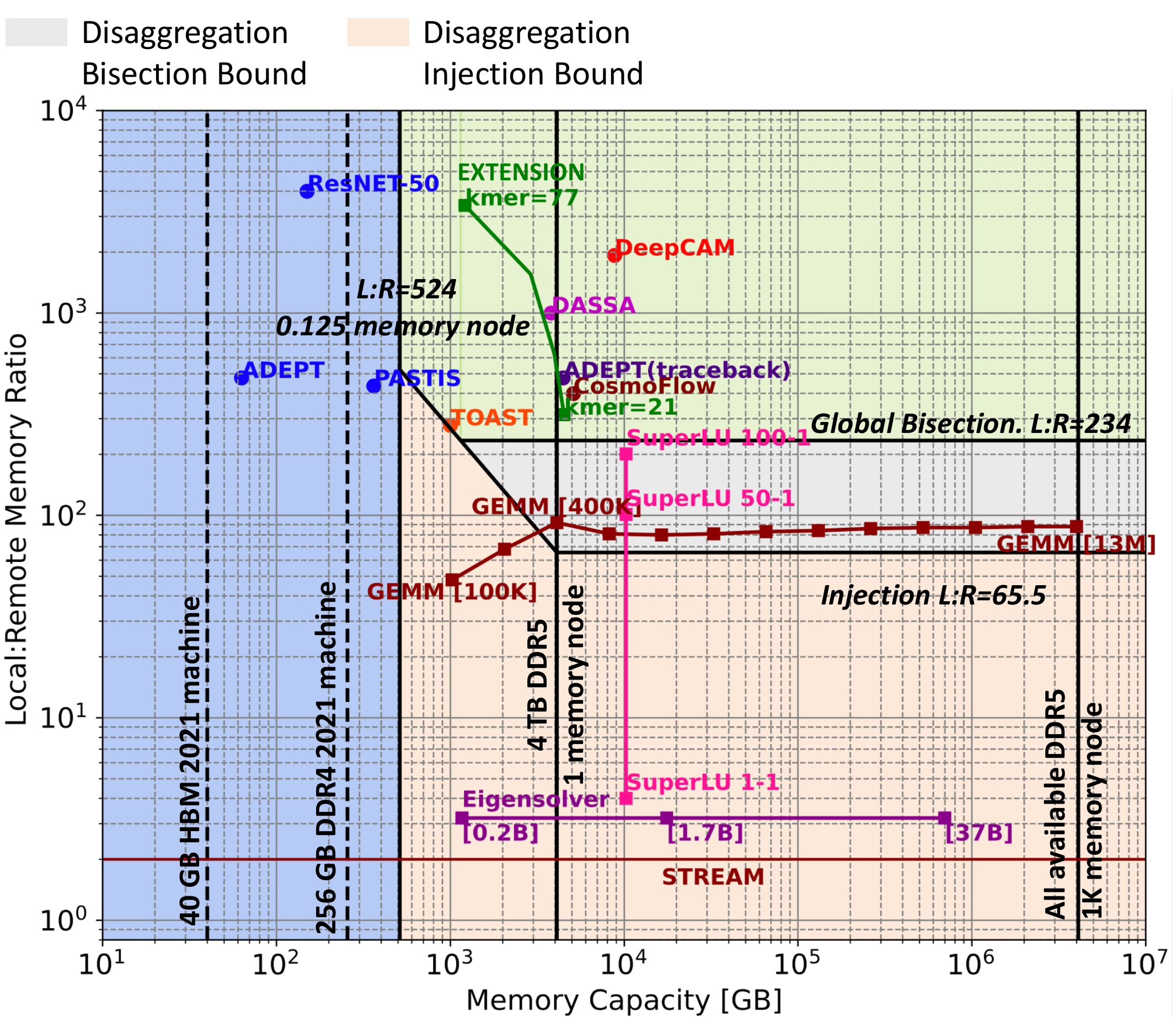}}
\caption{\label{fig:apps} Capacity and bottleneck visualization of applications on a disaggregated memory system. The L:R of PASTIS and ResNet-50 are profiled from Cori GPU~\cite{corigpu} at NERSC. ADEPT w/o traceback is modeled based on its implementation on Cori GPU. The corresponding matrix sizes $N\times N$ of GEMM and the eigensolver are marked in brackets as $[N]$.}
\end{minipage}
\end{figure}

Fig.~\ref{fig:app_metric_rack} and Fig.~\ref{fig:app_metric_global} visualizes a summary of all the tested applications in this section on a rack- and global- disaggregated memory system, respectively. Both figures combine the two critical metrics, the local and
remote memory access ratio (L:R) from Fig.~\ref{fig:memoryrooflineall} and the per-node memory capacity to provide an intuitive way to visualize the performance of applications on a future disaggregated memory system and assess individual application potential and pitfalls. Our system assumes 2026 memory technologies with each compute node having 512 GB HBM3 local memory, two times larger than a 2021 machine's node-local DDR capacity~\cite{perlmutter}.
Thus, applications that can fit in 2021 machine's node-local DDR can undoubtedly fit in future local memory. 
 We characterize the applications into four categories. 
 \textbf{Blue:~}required memory footprint can fit in local HBM memory. Thus, applications in this region would be HBM bound, e.g., ResNet-50. 
 \textbf{Green:~}required memory footprint does not fit in local memory, but applications could achieve HBM3 bandwidth due to a high L:R ratio (larger than 65.5) and would thus not incur a performance penalty from disaggregation, e.g., DeepCAM. Note applications in the green region are ultimately HBM bound but can still be impacted by the PCIe NIC bandwidth due to inefficient data movement and bandwidth contention, forcing them to fall into the orange zone. 
 \textbf{Orange:~}required memory footprint cannot fit in local memory, and they will be bound by the injection bandwidth due to the low L:R ratio (smaller than 65.5), e.g., STREAM ($>$512GB). 
\textbf{Grey:~}required memory footprint does not fit in local memory, and applications (65.5 $<$ L:R $<$ 234) will pay a performance penalty due to the bisection bandwidth, e.g., SuperLU. 
\textbf{Red:~}only for rack disaggregation. It represents taht there's not enough intra-rack remote memory.

The antidiagonal line connecting L:R=524 to L:R=65.5 in Fig.~\ref{fig:apps} shows the implications of network contention. If the memory capacity the application needs is between 512 GB and 4 TB, there are two design possibilities. The first is to use one memory node per compute node (L:R=65.5, one memory node), which guarantees all 100 GB/s of the PCIe6 NIC bandwidth but wastes memory capacity. The other possibility is to share memory nodes across compute nodes (upper left region in Fig.~\ref{fig:heatmap_capacity} and Fig.~\ref{fig:heatmap_pcie6}). In this case, memory is not wasted, but compute nodes must contend for memory node NIC bandwidth (L:R=524, 0.125 memory node). Such contention leads to an antidiagonal boundary between the green and orange zones. 
Following the same methodology, one could imagine disaggregating today's system can shrink the blue region to the 40GB HBM vertical dotted line. The green and orange zones expand to the left accordingly. The antidiagonal boundary moves to lower left, and the L:R boundary moves down a little to 62.2.

\textbf{ResNet-50:~}The ResNet-50 v1.5 is a 50-layer deep convolutional neural network. ResNet-50 has been implemented in both TensorFlow and PyTorch with numerous implementations and optimizations that prevent direct comparisons of system performance. The actual computation and memory characteristics of ResNet-50 come from Ibrahim~\etal~~\cite{ibrahim2021architectural}. ResNet-50 on Imagenet data requires 0.15 TB memory to store the training data set, and its L:R ratio is 3993. On the selected system configuration, the L:R ratio has no impact because the training data can easily fit into local memory.

\textbf{DeepCAM:~}
The DeepCAM climate benchmark is based on the 2018 work
of Kurth~\etal~\cite{kurth2018exascale} which was awarded the ACM Gordon Bell Prize. It uses deep learning to identify extreme weather phenomena from background images. Unlike ResNet-50, DeepCAM has a large memory requirement, 8.8 TB, to store the training size~\cite{ibrahim2021architectural}. It requires 2.2 memory nodes using our selected system configuration. As its L:R is 1927, which is higher than 65.5 (on the left of the orange dotted wall in Fig.~\ref{fig:memoryRoofline}), DeepCAM can operate at HBM3 speed on a disaggregated system that uses HBM3 as local memory and PCIe6 NIC for the network. 

\textbf{CosmoFlow:~}
CosmoFlow uses a 3D convolutional neural network with five convolutional layers and three fully connected layers. For the training run in Table~\ref{tab:ai_training_apps}, we can replicate the 5.1TB training data over 1.25 memory nodes per APU~\cite{ibrahim2021architectural} in the disaggregated memory system. It has an L:R of 399.
As the AI model size grows exponentially~\cite{aimodelsize}, it pushes the AI training workloads to have even larger memory requirements in the future. Therefore, AI training workloads with dense activation layers will result in a high L:R ratio and benefit from memory disaggregation (green zone). Alternatively, AI training workloads with shallow networks will pay the network bandwidth performance penalty on a disaggregated memory system (orange zone).

\textbf{DASSA:~}
The local similarity method is a time-domain data analysis algorithm developed to detect earthquakes in array seismic datasets~\cite{li2018high}. 
Each input file contains a 2D array (30,000 time samples and 11,648 channels).
For each cell in that 2D array, it calculates two correlations and each correlation needs to refer the other cells in a different channel in the window. With a typical window size of five hundred cells, one cell needs to access one thousand cells for its computation. Thus, the number of local memory accesses per cell is one thousand cells. The remote memory access is to stream the input data to the local memory once. As such, the number of remote memory accesses equals the total number of cells. This leads to an L:R ratio of 1,000.

\textbf{TOAST:~} The L:R ratio is calculated by profiling the DRAM data movement using the Intel VTune on one Cori Haswell~\cite{cori} node and dividing the input size. Thus, the L:R ratio is 278 and the required memory capacity is 1TB.

\textbf{ADEPT (no-traceback):~}
The core computation of ADEPT is to perform Smith-Waterman (SW) alignments~\cite{adept_awan}. SW is a dynamic programming algorithm that constructs an $m\times n$ matrix $A$ given two sequences of lengths $m$ and $n$. The matrix $A$ is used to find the optimal local alignment between the two sequences by listing all possible alignments. When operating in no-traceback mode, ADEPT is able to discard most of the matrix $A$ except the cells needed for the next iteration. When computing the matrix $A$, the score of any element $A(i, j)$ depends on elements $A(i, j-1)$, $A(i-1, j)$ and $A(i-1, j-1)$. The whole score matrix ($m\cdot n$) is maintained in the local memory. This leads to the number of local memory access to $3\cdot m\cdot n$, and the number of remote memory access to $m+n$. In this paper, we use a data set of about 31 million DNA reads and corresponding reference pairs with upper limits of $m=200$ and $n=780$ leading to the 63 GB total remote memory requirements with a L:R ratio of 477.

\textbf{ADEPT (traceback):~}
When operating in traceback mode, ADEPT kernel traces a path connecting the matrix cell with the highest score $(i,j), i>i_0,j> j_0$ back to the starting point ($i_0, j_0)$ in the matrix. For this, the full matrix must be available in the GPU's HBM therefore, this along with increasing total HBM requirements also requires additional memory accesses for traceback. The traceback step usually follows pointers starting from the highest scoring cell and ending when a cell with zero score is found, leading to additional $l$ memory accesses where $l$ represents the longest possible alignment. The longest alignment can at most be of the length $200$, i.e. longest possible read in the dataset. This also leads to an approximate L:R ratio of $477$.

\textbf{EXTENSION:~}
MetaHipMer's \cite{hofmeyr2020terabase} local assembly phase performs local extensions on sections of DNA with the help of DNA reads. 
We use the Arctic synth data set~\cite{mhm_steve} with four typical kmer size: 21, 33, 55 and 77 with 45 million extensions for four application runs. As such, one can observe that the L:R ratios vary from 314 to 3,402 with increasing kmer sizes.

\textbf{PASTIS:~}
We use a dataset that performs around 840 million pairwise alignments. This results in 158TB of local and 363GB of remote memory data movement (an L:R of 435).

\textbf{SuperLU:~} Fusion simulations use SuperLU as a preconditioner. It usually performs multiple iterations of sparse triangular solves per sparse LU factorization with one right-hand side. We define the sparsity of the input matrix as the number of non-zeros divided by the matrix size, and the matrix sparsity is 1e-3. The factored matrix (N$\times$ N) is a thousand times larger than the input matrix. Therefore, the memory requirement equals the total bytes of nonzeros in the LU factored matrix. Specifically, the number of nonzeros of the LU factored matrix in Figure~\ref{fig:apps} is 640 billion with N=25 million.
We apply the I/O model from Grigori et al.~\cite{grigori2010brief}to estimate the local data movement per local factorization using a 512GB HBM3 memory and a 40MB cache. For each local factorization (n $\times$ n), the data movement to/from cache is $\frac{n^{2/3}}{\sqrt{M}}$ word, where $M$ is the cache size and n equals 5.5million. The data movement in remote memory is to read the input matrix block for each local factorization and to write back the factored results. Thus the L:R for factorization ($LR_f$) equals to one. 
Each solve-iteration computes two triangular matrix-vector multiplications: $Lx=b$ and $Ux=b$. Triangular solves have two load operations for each non-zero (non-zero itself and the corresponding $b$) in $L$ and $U$ and a store to remote memory for the solution $x$. Thus, the L:R for one solve iteration is $\frac{nnz+n+2*nnz}{nnz+n} \approx 3$ as n $\ll$ nnz. The L:R for multiple solve iterations is $\frac{nnz+n+iter\times (2*nnz)}{nnz+n}$ which scales with the number of solve iterations per factorization. Following this method, the L:R for the entire SuperLU is 4, 101, and 201 with 1, 50, and 100 solve iterations per factorization. Thus, applications with 50 solve iterations per factorization will be bound by rack bisection as Fig.~\ref{fig:app_metric_rack} shows. Alternatively, applications with 100 solve iterations per factorization will pay global bisection as presented in Fig.~\ref{fig:app_metric_global}.

\textbf{Eigensolver:~}
The MFDn eigensolver performs sparse matrix-matrix multiplication (SpMM). We refer to the I/O model for SpMM~\cite{bender2007optimal} to estimate our local memory data movement, $ (k\times N)\times (1 + \log_M ((k\times N)/M))$, where $N$ is the matrix dimension, $k \times N$ gives the total number of non-zeros, and $M$ is the cache size. Again, we consider the cache size is 40MB. The remote memory data movement (R) reads the input matrix and stores the results. Thus, it gives a constant L:R ratio of 3.2 as the matrix size varies from 0.2 billion by 0.2 billion to 37 billion by 37 billion, and the sparsity (total number of nonzeros divided by the matrix size) varies from 1e-6 to 1e-7. Since the input matrix is symmetric, the memory requirement equals the size of half the number of nonzeros in the input matrix.

\textbf{GEMM:~} The general matrix multiplication (GEMM) kernel is defined as the operation $C = A\cdot B$, with $A$ and $B$ as matrix inputs and $C$ as the output. We assume all three matrices are square double-precision matrices (N $\times$ N) with $N$ being the maximum dimension that fits in DDR memory. Using estimates derived from the Holder-Brascamp-Leib (HBL) inequality~\cite{smith2017tight}, we may estimate the data movement ($R$) to/from remote memory as $\frac{2\cdot N^{3}}{\sqrt M}+N^2-3\cdot M$ elements where $M$ is the local memory (cache) capacity in elements (64G).
We apply this model recursively to estimate the local data movement per local GEMM using a 512GB memory and a 40MB cache~\cite{nvidia2020a100}. This number is scaled by the requisite number of local GEMMs ($(\frac{DDR}{HBM})^\frac{3}{2}$) to produce the L:R ratio which we observe varies from about 50 to 90. It is worth noticing that a bigger GEMM, e.g., matrix size is 400K$\times$400K, can help to eliminate the injection bandwidth penalty. However, GEMM will ultimately pay the rack bisection bandwidth penalty (Fig.~\ref{fig:app_metric_rack}) because its L:R  remains close to 90 no matter how big the matrix is.

\textbf{STREAM:~}STREAM TRIAD is a benchmark that measures sustainable memory bandwidth.
It computes a vector operation $C(i) = A(i) + \alpha \cdot B(i)$. This operation involves two loads ($A(i)$ and $B(i)$) and one store ($C(i)$) in the remote memory. Reads from (writes to) remote memory incur writes(reads) in local memory on top of nominal reads/writes in local memory. Thus, the L:R equals 2.

\subsection{Application Analysis Summary}
The case studies represent a diverse array of memory access patterns and memory capacity needs across multiple domains. This trend will likely continue as scientific workloads evolve over time. 

For the exemplar disaggregated memory system configuration consisting of 10,000 compute nodes and 1,000 memory nodes, we see that nine out of thirteen workloads fall into the blue and green zones, which will not suffer penalties from bisection bandwidth. The SuperLU\_DIST with 100 solves per factorization pays global bisection but is not sensitive to rack bisection. 
Only the STREAM and the Eigensolver fall into the orange zone and could see a penalty from injection bandwidth. 

Although we proxied future L:R ratios using today's problem sizes, we believe future L:R ratios will be at least as big as surface to volume ratios never shrink. 
Ultimately, these applications are unlikely to see any performance loss from disaggregated memory over the existing state of the practice.

Assuming these applications represent a workload and the applications falling into the green and orange zones constitute less than 10\% of the total workload node hours, then the disaggregate system discussed here could result in significant cost savings, eliminating 40PB of node-local DDR memory (10K compute nodes$\times$4TB) of memory in exchange of 1000 memory nodes of 4TB each without hurting performance.

\section{Discussion and Conclusions}\label{sec5}
In this paper, we focused on architecture, bottlenecks, and characterization of applications running on disaggregated memory system architectures in the 2024-2026 time frame. As visualized in Fig.~\ref{fig:apps}, 9 out of 13 of the applications we examined either have sufficiently low memory requirements that they can comfortably fit in a future APU's HBM memory or have a sufficiently high local:remote data movement ratio that the architected local:remote bandwidth tapering will not impede performance. Nevertheless, it is imperative HPC system architects and vendors follow a few design principles lest the potential remote bandwidth be underutilized.

{\bf Rack- and Global Disaggregation:~} System architects need to decide whether to do intra-rack disaggregation or system-wide disaggregation. Our examined applications in Fig.~\ref{fig:apps} suggest that intra-rack disaggregation can meet the applications' memory requirement and provide sufficient remote memory bandwidth. It also avoids the increased memory controller overhead of full-system disaggregation~\cite{zervas2018optically,michelogiannakis2023efficient}. 
However, the impact of memory controller overhead still needs to be explored.

{\bf Memory Extension:~}System architects must decide whether remote memory is exposed as a second NUMA node with data movement affected via RDMA (e.g. SHMEM put/get) or uncacheable load/store instructions -- or -- whether HBM should be viewed as either a hardware-controlled line cache or OS-controlled page cache. The nuance arises in whether applications and processors can express concurrency greater than remote memory's latency-bandwidth product~\cite{LittleLaw} given a latency comparable to the 2us observed on a 2021 HPC system and bandwidth varying from PCIe4's 25GB/s to PCIe6's 100GB/s.

\begin{figure}[htb]
\centering
\includegraphics[width=0.6\linewidth]{ 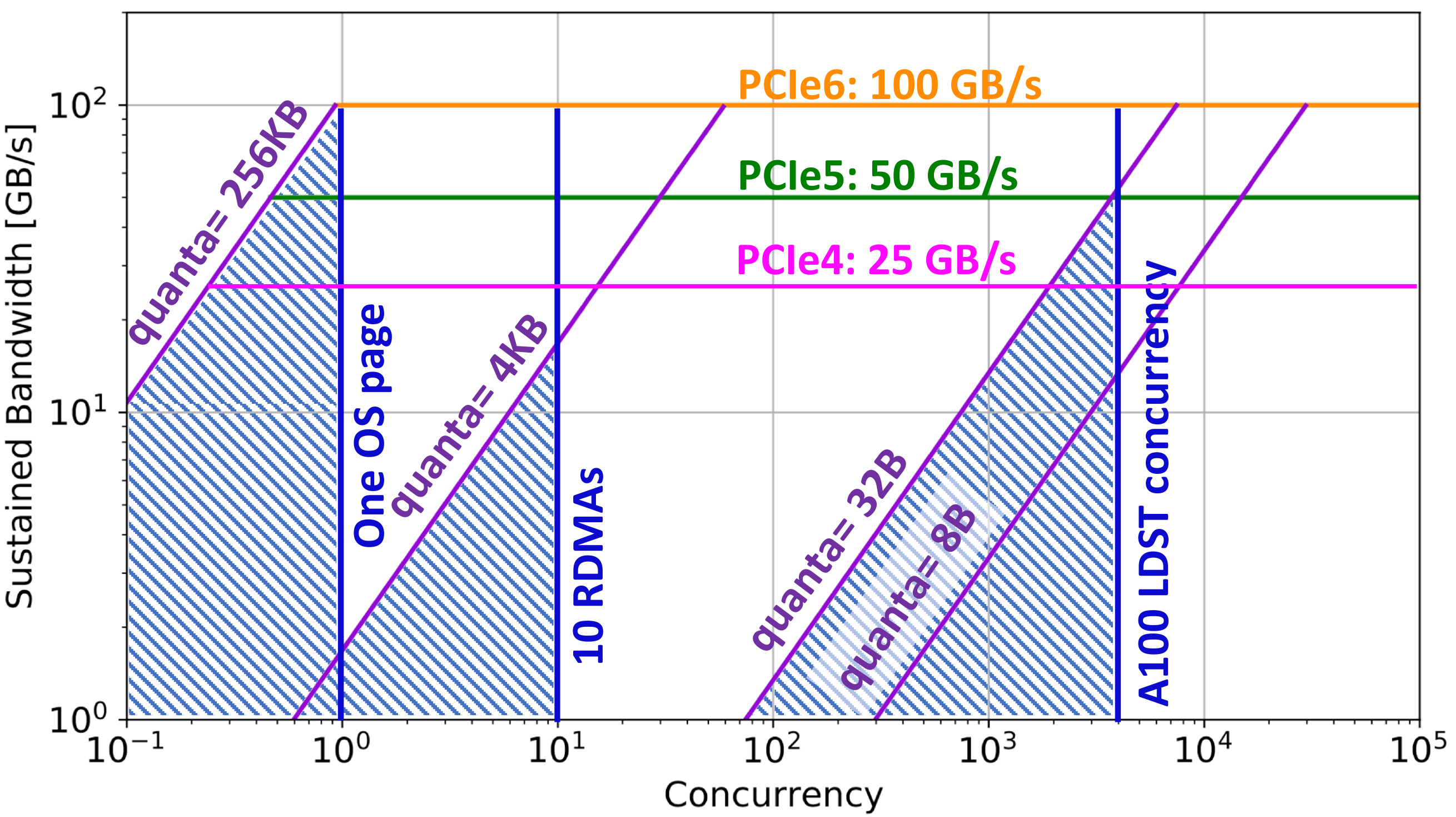}
\caption{\label{fig:solutions} The Roofline-inspired visualization of Little's Law (Concurrency Roofline) informs architects of viable combinations of network bandwidth, memory access quanta, and concurrency. When the intercept of quanta and concurrency is less than a PCIe bandwidth, sustained bandwidth is impeded (e.g. 4KB OS paging or 32B caching on the A100). 
}
\end{figure}

Inspired by the Roofline model~\cite{williams2009Roofline}, Figure~\ref{fig:solutions} plots the impact of Little's Law on memory bandwidth for varying access quanta (diagonals) and concurrency (vertical lines). System architects must choose a quanta that attains available bandwidth at an application concurrency less than the processor/system concurrency upper bound. For example, an OS cache sustaining only one outstanding page fault (concurrency$\le$1) will never be able to sustain even PCIe4 bandwidth with 4KB pages. Similarly, an A100 GPU has insufficient load/store concurrency to sustain PCIe5 bandwidth using coalesced 32B cache lines. Ultimately, vendors must provide either larger pages (e.g $\ge$256KB), $\ge$64B cache lines, twice as many load/store units as an A100 GPU, or demand applications continually initiate hundreds of KB-sized asynchronous RDMAs (spread across multiple processes).

{\bf Lustre Replacement:~}Lustre is superfluous for applications requiring either private or read-only file access (e.g. AI training data). Vendors wishing to leverage remote memory nodes as a replacement for distributed file systems must provide a file system interface and guarantee durability for the life of a job (more than an individual executable).
Our Roofline-inspired Little's Law analysis (Fig.~\ref{fig:solutions}) still applies. That is, assuming file system software overhead is far less than network latency, applications must both continually read/write 256KB blocks in order to sustain PCIe6 bandwidths whilst ensuring the local to file I/O data movement ratio exceeds 65.

{\bf Inter-Process Communication:~}Whereas memory and inter-process communication (e.g. MPI) were traditionally architected with dedicated bandwidths, in disaggregated systems, they will contend for finite PCIe bandwidth. Moreover, collective and point-to-point communication will also contend for the bisection and inject bandwidth. As such,
applications with even a modest inter-process to remote memory ratio may see PCIe bandwidth emerge as a bottleneck. Applications with heavy collective communications may be sensitive to the bisection bandwidth.

{\bf Future Portents:~}Historically, latency lags bandwidth~\cite{patterson2004latency}, and to a lesser degree, one expects remote bandwidth to lag local bandwidth. As such, beyond 2026 we expect the latency-bandwidth product (requisite concurrency) to increase nearly as fast as remote bandwidth. Systems in that time frame with require even larger pages, even more concurrent RDMAs (easily realized with more processes per node), or GPUs with even more concurrency (almost guaranteed). Similarly, the hardware local:remote ratio will increase slowly, implying some applications may become remote memory-limited. As such, memory disaggregation will likely continue to be a viable approach so long as network bandwidth increases.

{\bf Workload Analysis:~}Whereas this paper focused on analyzing individual applications in a system with disaggregated memory, the ultimate efficacy of such a system is premised on the specific workload requirements. Practitioners wishing to leverage our methodology should characterize their applications along the lines of Fig.~\ref{fig:apps} and ratio their compute to memory nodes as: the sum of the node hours of all the applications falling into the blue region divided by the sum of the node hours of all the applications falling into the green and orange regions (scaled by memory capacity/4TB). If the scaled node hours of the green and orange regions dominate, then it indicates that as such a ratio will demand more memory nodes than compute nodes.
Similarly, if the scaled node hours of the orange region dominate, the workload is better served with node-local DDR unencumbered by limited PCIe bandwidths.

For amenable workloads and collaborative vendors, memory disaggregation will provide a cost-effective means of mitigating the dynamic and highly variable memory requirements found in HPC centers.

\section*{Acknowledgments}
This material is based upon work supported by the Advanced Scientific Computing Research Program in the U.S. Department of Energy, Office of Science, under Award Number DE-AC02-05CH11231 and used resources of the National Energy Research Scientific Computing Center (NERSC), which is supported by the Office of Science of the U.S. Department of Energy under Contract No. DE-AC02-05CH11231. Pieter Maris is supported by the US Department of Energy, Office of Science, under Grant DE-SC0023495. Khaled Ibrahim and Tan Nguyen were especially helpful in answering questions on AI workloads. John Wu and Bin Dong were very helpful in providing their knowledge on DASSA.

\bibliography{wileyNJD-AMA}

\end{document}